\documentclass[5p,twocolumn]{elsarticle}

\usepackage{graphicx}
\usepackage{dcolumn}
\usepackage{pifont}
\usepackage{bm}
\usepackage{multirow}
\usepackage{amsmath}
\usepackage{float}
\usepackage{txfonts}
\usepackage[version=3]{mhchem}

\usepackage[usenames,dvipsnames]{xcolor}
\definecolor{blue}{RGB}{0,0,225}
\definecolor{cream}{RGB}{222,217,201}
\definecolor{red}{RGB}{225,0,0}

\journal{Elsevier}

\begin{document}
\title{Influence of M/A substitution on material properties of intermetallic compounds MSn$_2$ ~(M = Fe, Co; A = Li, Na): A first-principles study}

\author[kimuniv-m]{Chol-Jun Yu\corref{cor}}
\cortext[cor]{Corresponding author}
\ead{cj.yu@ryongnamsan.edu.kp}
\author[kimuniv-m]{Un-Song Hwang}
\author[kimuniv-m]{Yong-Chol Pak}
\author[kimuniv-m]{Kyonga Rim}
\author[kimuniv-m]{Chol Ryu}
\author[kimuniv-m]{Chon-Ryong Mun}
\author[kimuniv-m,kimuniv-n]{Un-Gi Jong}

\address[kimuniv-m]{Chair of Computational Materials Design, Faculty of Materials Science, Kim Il Sung University, Ryongnam-Dong, Taesong District, Pyongyang, Democratic People's Republic of Korea}
\address[kimuniv-n]{Natural Science Center, Kim Il Sung University, Ryongnam-Dong, Taesong District, Pyongyang, Democratic People's Republic of Korea}

\begin{abstract}
Iron and cobalt distannides \ce{MSn2} (M = Fe, Co) are regarded as a promising conversion-type anode material for lithium- and sodium-ion batteries, but their properties are not well understood.
In this work, we report a first-principles study of alkali metal (A = Li, Na) substitutional effect on the structural, mechanical, lattice vibrational, electronic and defect properties of these distannides.
Special attention is paid to systematic comparison between \ce{FeSn2} and \ce{CoSn2}.
Our calculations reveal that M/A substitution induces a lattice expansion and decrease of elastic constants, which is more announced with Na substitution than Li, and moreover changes the elastic property of \ce{FeSn2} from ductile to brittle whereas preserves the ductility of \ce{CoSn2}.
An imaginary phonon frequency mode appears only for \ce{FeSn2} and \ce{FeNaSn2}, and M/A substitution provokes a definite gap between high and low frequency regions.
We perform a careful analysis of electronic density of states, band structures and Fermi surface, providing an insight into difference of electronic structures between \ce{FeSn2} and \ce{CoSn2}.
With further calculation of defect formation energies and alkali ion diffusion barriers, we believe this work can be useful to design conversion-type anode materials for alkali-ion batteries.
\end{abstract}

\maketitle

\section{Introduction}
Alkali (lithium and sodium) ion batteries have been attracting considerable attention as a promising power source of electronic devices and electric vehicles due to their high energy density and long cycle life.
When compared with lithium, sodium has a merit of resource abundance on the Earth's crust, leading to apparent cost lowering of battery production and thus enabling sodium ion batteries (SIBs) to be commercially viable for electric vehicles and stationary energy storage~\cite{Vaalma18nrm}.
However, no optimal electrode materials have yet been developed as some scientific challenges unresolved~\cite{Mukherjee19m,Bai17aem}.
In particular, since graphite, the commercial intercalation-type anode material for lithium ion batteries (LIBs), exhibits extremely low specific capacity when reacting with sodium, numerous works have been devoted to finding suitable anode materials for SIBs~\cite{Luo16acr}.

The key issues in development of high performance anode materials can be summarized as high reversible capacity, long cycle life, and high rate capability.
With these respects, several kinds of anode materials with their own merits and demerits have been found so far, including carbon- and alloy-based materials, metal oxides and 2D materials~\cite{Li18ees}.
Among these, alloy-based materials have earned a remarkable interest because of their very high capacity and low redox potential, in spite of their critical problem of poor cycling stability.
Typically, metallic $\beta$-Sn is a conversion-type anode material for both LIBs and SIBs, exhibiting maximum theoretical capacities of 994 mAh/g for complete conversion of \ce{Li22Sn5}~\cite{Huang18jps} and 847 mAh/g for \ce{Na15Sn4}~\cite{Stratford17jacs,Li15acr,Baggetto14jmca,Wang12nl}.
However, the application of $\beta$-Sn as anode material is limited by its fast fading of performance due to the low electronic conductivity of Sn and in particular large volume changes of $\sim$300\% for Li and 420\% for Na during the charge/discharge processes, leading to crack and pulverization of Sn particles and depletion of electrolyte.

As one way to mitigate such problems, active $\beta$-Sn has been proposed to be alloyed with inactive transition metals that do not directly react with lithium and sodium during cycling~\cite{Wang19jac,Walter16jmca,Zhang17ra,Wang10aami}.
The transition metals were known to lessen the volume change by forming a buffer framework, enhance the electric conductivity, limit the coalescence and increase the dispersion of Li$-$ or Na$-$Sn particles~\cite{Wang10aami,Chamas13cm}.
Accordingly, iron and cobalt have already been used with tin to form \ce{FeSn2} and \ce{CoSn2} alloys as promising anode materials with high capacity and stable cycling performance for SIBs~\cite{Wang18n,Vogt17jmca,Vogt16jes,Yui15jes,Gonzalez13cec} as well as LIBs~\cite{Chamas11ea,Chamas11jps,Zhang08jac}.
Recently, Vogt and Villevieille~\cite{Vogt17jmca,Vogt16jes} have reported that these alloys can achieve high capacities of up to 680 mAh/g when fully converted to \ce{Na15Sn4} (twice larger than hard carbon), demonstrating the significant influence of inactive metals on the reaction mechanism.
In addition, their composites with carbon~\cite{Edison17jps,Edison17cec,Liu15ssi,Nacimiento12us} and other materials~\cite{Leibowitz15jps,Guo07ea} have been found to exhibit highly stable electrochemical performance as anodes for both LIBs and SIBs.
Despite such extensive experimental studies, no theoretical study of \ce{FeSn2} and \ce{CoSn2} in terms of Li or Na reaction has yet reported, except some first-principles studies of Fe$-$ and Co$-$Sn alloys~\cite{Sun16cms,Jong15sd}

In this work, we have investigated the electronic structures, lattice vibrational and elastic properties, and point defect energetics of \ce{FeSn2} and \ce{CoSn2} as reacting with Li and Na, by using first-principles calculations within the density functional theory (DFT) framework.
We first determined the lowest energy spin configurations of \ce{FeSn2} and \ce{CoSn2} by conducting structural optimizations, and using these structures, calculated their electronic band structures with partial density of states (PDOS), elastic constants and phonon dispersion curves.
Then we have investigated their chemical reactivity with sodium and lithium by estimating the formation energies of point defects including vacancies and antisites in bulk.
Systematic comparison between Fe and Co was provided in each calculation.

\section{Theoretical methods\label{sec_meth}}
\subsection{Structural models}
\begin{figure}[!th]
\centering
\includegraphics[clip=true,scale=0.1]{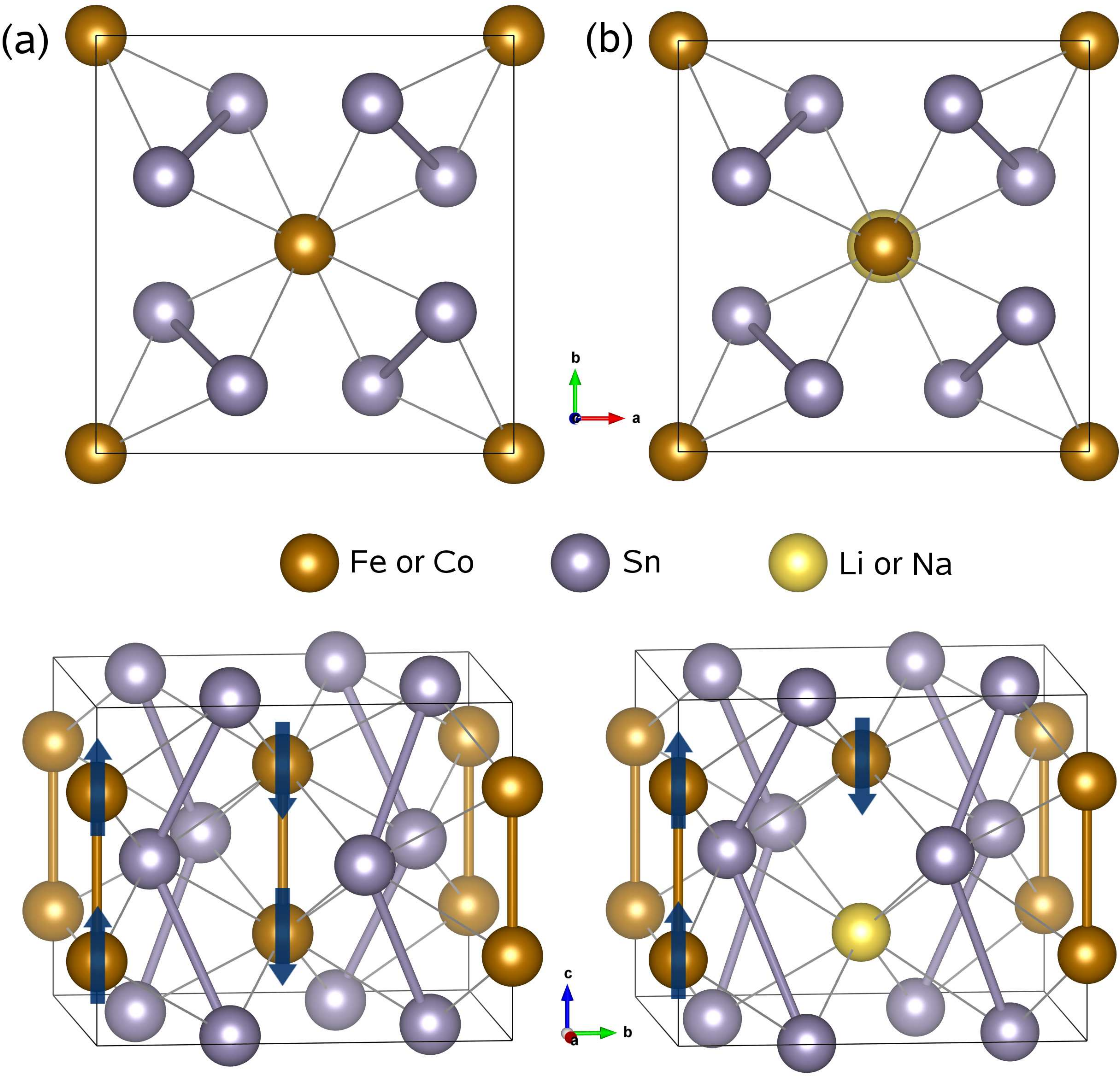}
\caption{Crystalline structure of conventional unit cell for (a) \ce{MSn2} (M = Fe, Co) with four formula units and (b) \ce{M_{3/4}A_{1/4}Sn2} (A = Li, Na). Upper and lower panels show top and perspective views respectively. Blue-colored arrows indicate the spin direction of transition metal atoms, indicating the anti-ferromagnetic (AFM) configuration here.}
\label{fig1}
\end{figure}

The compounds \ce{FeSn2} and \ce{CoSn2} are known to crystallize in the \ce{CuAl2}-type tetragonal structure with a space group of $I4/mcm$~\cite{Armbruster07zkn}.
As shown in Fig.~\ref{fig1}(a), there are four formula units (12 atoms) in the conventional unit cell, where transition metal atoms occupy the $4a$ (0, 0, 0.25) sites and Sn atoms locate at the $8h$ ($x$, 0.5+$x$, 0) sites ($x=0.1611$ in \ce{FeSn2} and $0.1649$ in \ce{CoSn2})~\cite{Armbruster07zkn}.
In order to study the influence of reaction with Li and Na on their properties, we consider M/A exchange structural models (M = Fe, Co; A = Li, Na), which were constructed by replacing one Fe or Co atom with Li or Na atom in the conventional unit cell, leading to formation of tetragonal \ce{M_{3/4}A_{1/4}Sn2} (or simply \ce{MASn2}) compound with a space group of $P422$, as shown in Fig.~\ref{fig1}(b).
Also we constructed the $2\times2\times2$ supercells (96 atoms) to study lattice vibration properties and furthermore point defects including vacancies ($V_{\text{M}}$ and $V_{\text{Sn}}$) and substitutions (A$_{\text{M}}$ and A$_{\text{Sn}}$).

\subsection{Computational details}
All the DFT calculations have been carried out by applying the pseudopotential plane wave method as implemented in {\footnotesize QUANTUM ESPRESSO} (QE, version 6.2) package~\cite{QE09jpcm}.
For a description of the Coulombic interaction between the ionic cores and the valence electrons, we have constructed the ultrasoft pseudopotentials of the atoms by executing LD1 code included in the QE package, using the input files provided in the PS library (1.0).
The valence electron configurations of atoms are Na: $2s^22p^63s^1$, Li: $1s^22s^12p^0$, Fe: $3s^23p^63d^{6}4s^2$, Co: $3s^23p^63d^{7}4s^2$, and Sn: $4d^{10}5s^25p^2$.
The Perdew-Burke-Ernzerhof (PBE) formulation~\cite{PBE96prl} within generalized gradient approximation (GGA) was adopted to describe the exchange-correlation interaction among the valence electrons.

Structural optimizations of the conventional unit cells containing 4 formula units were carried out with the kinetic cutoff energies of 60 Ry for wave function and 600 Ry for electron density, and the special $k$-points with a $(8\times 8\times 10)$ mesh.
The atoms were relaxed until the forces converged to $5\times 10^{-4}$ Ry/Bohr, while the crystalline lattices were allowed to vary until the pressure became less than 0.005 GPa.
To achieve a convergence of self-consistent cycle, the Methfessel-Paxton first-order smearing method was applied with a gaussian spreading parameter of 0.02 Ry.  
For the calculation of energy band structures and density of states (DOS), we used a denser $k$-point mesh of $(10\times 10\times 12)$ with a tetrahedron occupation.
In these calculations, spin-polarization effect was considered by applying three different magnetic orderings such as ferromagnetic (FM), anti-ferromagnetic (AFM) and non-magnetic (NM) states.
The phonon dispersion curves and the corresponding phonon DOS were calculated using the finite displacement method, as implemented in {\footnotesize Phonopy} code~\cite{phonopy}.
The $2\times2\times2$ supercells were adopted with reduced $k$-point sampling of ($2\times2\times2$) in accordance to the larger size of supercell, while the ($30\times30\times30$) $q$-point mesh was used for the phonon DOS calculation.
The six independent elastic stiffness constants were determined based on the efficient stress-strain method, in which a set of homogeneous deformations with maximum strain amplitude of 0.005 GPa were applied and the resulting stress with respect to the internal degrees of freedom was calculated.
Once obtained the elastic stiffness matrix, the elastic compliance constants were also evaluated by calculating the inverse matrix.

In the defect calculations, all the atoms were relaxed while fixing the lattice constants, with reduced computational parameters such as cutoff energy of 30 Ry and $k$-point mesh of ($2\times2\times2$).
Moreover, we determined the activation barriers for Li and Na atom diffusion along the vacancy-mediated paths by applying the climbing image nudged elastic band (NEB) method~\cite{NEB00jcp}.
During the NEB run, the supercell sizes were fixed at the optimized ones, and all the atoms were allowed to relax.
The number of NEB image points was seven, and the convergence threshold for force on the elastic band was 0.05 eV/\AA.
Visualization of crystalline lattice and volumetric data of charge density was performed by using the {\footnotesize VESTA} code~\cite{VESTA11jac}.

\subsection{Theory}
The mechanical properties of a polycrystalline solid are estimated by elastic moduli such as bulk ($B$), shear ($G$) and Young's ($E$) moduli.
These can be determined from the set of elastic stiffness ($C_{ij}$) and compliance ($S_{ij}$) constants calculated for a single crystal.
There are six independent elastic constants for tetragonal structure: $ij=11, 12, 13, 33, 44$ and $66$.
The bulk and shear moduli can be determined using the stiffness constants within the Voigt approximation as follows,
\begin{equation}
\begin{gathered}
B_{\text{V}}=\frac{2C_{11}+C_{33}+2(C_{12}+2C_{13})}{9}, \\
G_{\text{V}}=\frac{2C_{11}+C_{33}-(C_{12}+2C_{13})+3(2C_{44}+C_{66})}{15}
\end{gathered}
\end{equation}
Meanwhile, they can also be calculated using the compliance constants within the Reuss approximation as follows,
\begin{equation}
\begin{gathered}
B_{\text{R}}=\frac{1}{2S_{11}+S_{33}+2(S_{12}+2S_{13})}, \\
G_{\text{R}}=\frac{15}{4(2S_{11}+S_{33})-4(S_{12}+2S_{13})+3(2S_{44}+S_{66})}
\end{gathered}
\end{equation}
As indicated by Hill, the Voigt and Reuss approaches yield the lower and upper limits of the polycrystalline moduli and the real moduli are estimated by arithmetic mean value as follows,
\begin{equation}
B=\frac{B_{\text{V}}+B_{\text{R}}}{2}, 
G=\frac{G_{\text{V}}+G_{\text{R}}}{2}
\end{equation}
Then, the Young's modulus and Poisson's ratio ($\nu$) are evaluated from the calculated bulk and shear moduli as follows,
\begin{equation}
E=\frac{9BG}{3B+G}, 
\nu=\frac{3B-2G}{6B+2G}
\end{equation}

The elastic moduli can be used to determine the longitudinal ($\upsilon_l$) and transverse ($\upsilon_t$) elastic wave velocities as follows,
\begin{equation}
\upsilon_l=\sqrt{\frac{3B+4G}{3\rho}},
\upsilon_t=\sqrt{\frac{G}{\rho}}
\end{equation}
where $\rho$ is the density.
Using these values, the average sound velocity $\upsilon_m$ is evaluated as follows,
\begin{equation}
\upsilon_m=\left[\frac{1}{3}\left(\frac{2}{\upsilon_t^3}+\frac{1}{\upsilon_l^3}\right)\right]^{-1/3}
\end{equation}
Then, as an important thermodynamic parameter for checking the degree of mechanical properties, the Debye temperature $\theta_{\text{D}}$ is estimated using the following equation,
\begin{equation}
\theta_{\text{D}}=\frac{h}{k_{\text{B}}}\left[\frac{3N}{4\pi V}\right]^{1/3}\upsilon_m
\end{equation}
where $h$ and $k_{\text{B}}$ are the Plank's and Boltzmann's constants, $N$ the number of atoms in the unit cell and $V$ the unit cell volume.

The formation energy of point defect $D$ is calculated using the total energies as follows,
\begin{equation}
E_f=E[D]-E_{\text{perf}}-\sum_in_i\mu_i \label{eq_df}
\end{equation}
where $E[D]$ and $E_{\text{perf}}$ are the total energy of compounds with and without defect, and $n_i$ is the number of added ($n_i>0$) or removed ($n_i<0$) $i$-type species with a chemical potential of $\mu_i$.
The chemical potential of metal atoms were estimated as the total energy per atom in their bulk systems: $bcc$-Fe, $hcp$-Co, diamond cubic-Sn, and $bcc$-Li or -Na.
To check the thermodynamic stability of the compound with defect, its formation enthalpy per formula unit was calculated as follows,
\begin{equation}
\Delta H_{\text{M}_m\text{A}_n\text{Sn}_2}=\frac{E_{\text{M}_m\text{A}_n\text{Sn}_2}-mE_{\text{M}}-nE_{\text{A}}-2E_{\text{Sn}}}{m+n+2} \label{eq_hf}
\end{equation}
where $E_{\text{M}_m\text{A}_n\text{Sn}_2}$, $E_{\text{M}}$, $E_{\text{A}}$ and $E_{\text{Sn}}$ are the total energy of \ce{M_{$m$}A_{$n$}Sn2} compound, pure transition, alkali and Sn metals, and $m$ and $n$ are the corresponding number of atoms in the unit cell.

\begin{table*}[!t]
\small
\caption{Optimized lattice constants $a$ and $c$, tetragonal ratio $c/a$, unit cell volume $V$, Sn position $x_{\text{Sn}}$, mass density $\rho$, M$-$M interatomic distance $d_{\ce{M}-\ce{M}}$, cohesive energy per atom $E_c$ and formation energy per atom $E_f$ of intermetallics.}
\label{tabl1}
\begin{tabular}{lccccccccc}
\hline
Compound & $a$ (\AA) & $c$ (\AA) & $c/a$ & $V$ (\AA$^3$) & $x_{\text{Sn}}$ & $\rho$ (g/cm$^3$) & $d_{\ce{M}-\ce{M}}$ (\AA) & $E_c$ (eV) & $E_f$ (eV) \\
\hline
\ce{FeSn2} & 6.5333 & 5.3271 & 0.8153 & 227.38 & 0.1623 & 8.566 & 2.663 & $-5.155$ & $-0.221$ \\
\ce{FeSn2}$^a$ & 6.5331 & 5.3202 & 0.8144 & 227.07 & 0.1611 & 8.577 & & & ~$-0.148^b$ \\
\ce{Fe_{3/4}Li_{1/4}Sn2} & 6.5895 & 5.4728 & 0.8305 & 237.64 & 0.1628 & 7.854 & 2.736 & $-4.650$ & $-0.164$ \\
\ce{Fe_{3/4}Na_{1/4}Sn2} & 6.6831 & 5.6733 & 0.8489 & 253.39 & 0.1539 & 7.471 & 2.837 & $-4.476$ & $-0.035$ \\
\hline
\ce{CoSn2} & 6.3442 & 5.4723 & 0.8626 & 220.25 & 0.1669 & 8.936 & 2.736 & $-4.835$ & $-0.183$ \\
\ce{CoSn2}$^a$ & 6.3617 & 5.4582 & 0.8580 & 220.90 & 0.1649 & 8.910 & & & ~$-0.177^b$ \\
\ce{Co_{3/4}Li_{1/4}Sn2} & 6.4781 & 5.5963 & 0.8639 & 234.86 & 0.1608 & 8.013 & 2.798 & $-4.416$ & $-0.141$ \\
\ce{Co_{3/4}Na_{1/4}Sn2} & 6.5733 & 5.7381 & 0.8729 & 247.93 & 0.1520 & 7.698 & 2.869 & $-4.244$ & $-0.015$ \\
\hline
\end{tabular} \\
$^a$ X-ray diffraction data at $T$ = 295 K~\cite{Armbruster07zkn}. \\
$^b$ DFT calculation data with PBE functional~\cite{Sun16cms}
\end{table*}

\section{Results and discussion}
\subsection{Structural properties with magnetic ordering}
Firstly we determined the favorable spin configuration for transition metal atoms in \ce{MSn2} and \ce{MASn2} with crystalline lattice optimization.
It was found that for the case of \ce{FeSn2} the AFM configuration was energetically favorable with the best agreement of lattice constants to the experiment (see Table S1 in Supplementary Information).
In this AFM state, the Fe atoms have the magnetic moment of $\pm$1.89 $\mu_B$, while the total magnetization was confirmed to be zero (see Table S2).
For the case of \ce{CoSn2}, however, the NM state was always observed though the three different spin configurations of AFM, FM and NM were initially imposed.
In fact, Co atoms with these different initial impositions were found to have zero magnetic moment after lattice optimization and SCF cycle.
When replacing M (Fe, Co) atom by A (Li, Na) atom, although we initially imposed the AFM configurations on the resultant unit cell of \ce{MASn2}, the FM state was observed for \ce{FeASn2} while also the NM state was realized for \ce{CoASn2} (see Table S2).

\begin{figure}[!th]
\centering
\includegraphics[clip=true,scale=0.5]{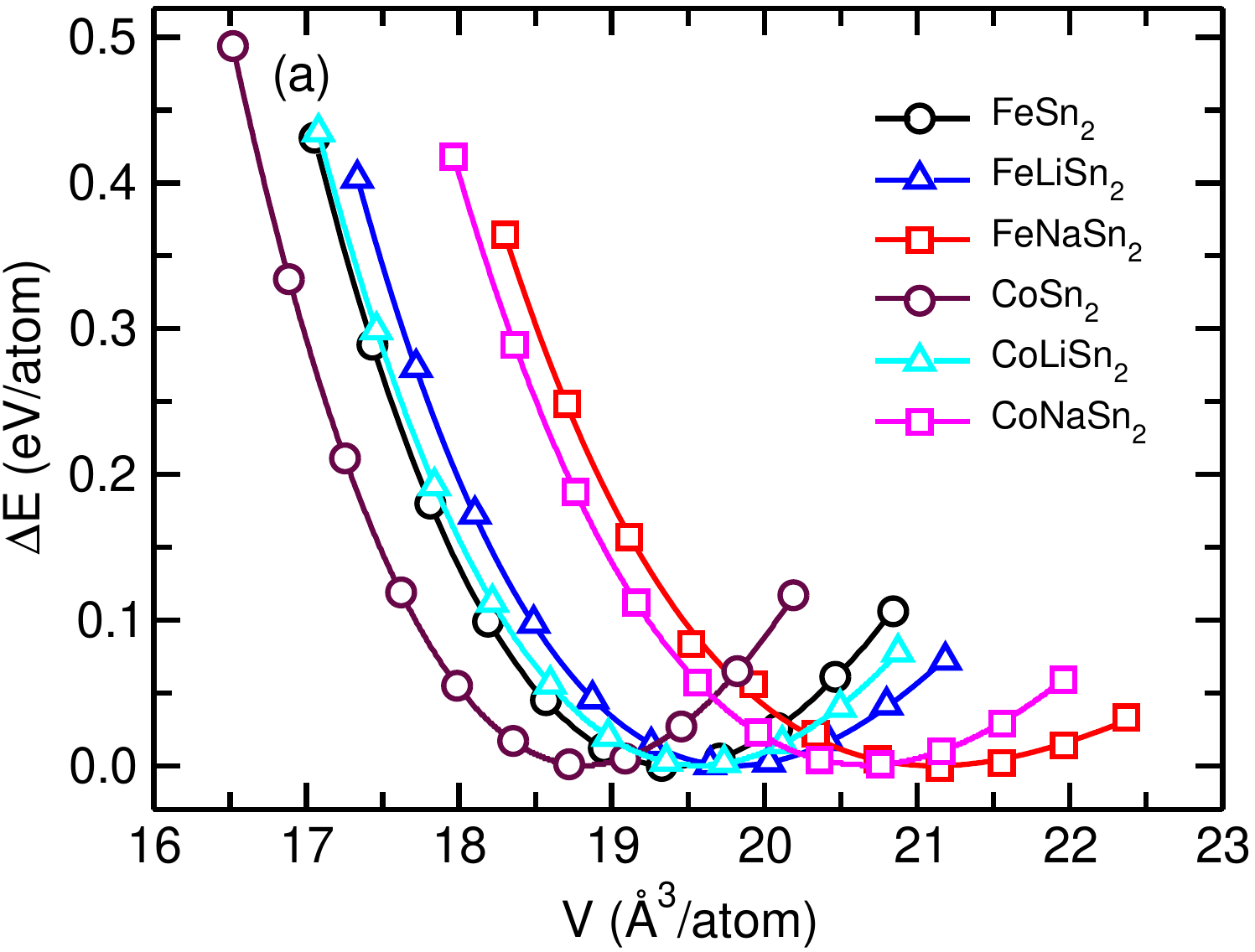}
\includegraphics[clip=true,scale=0.5]{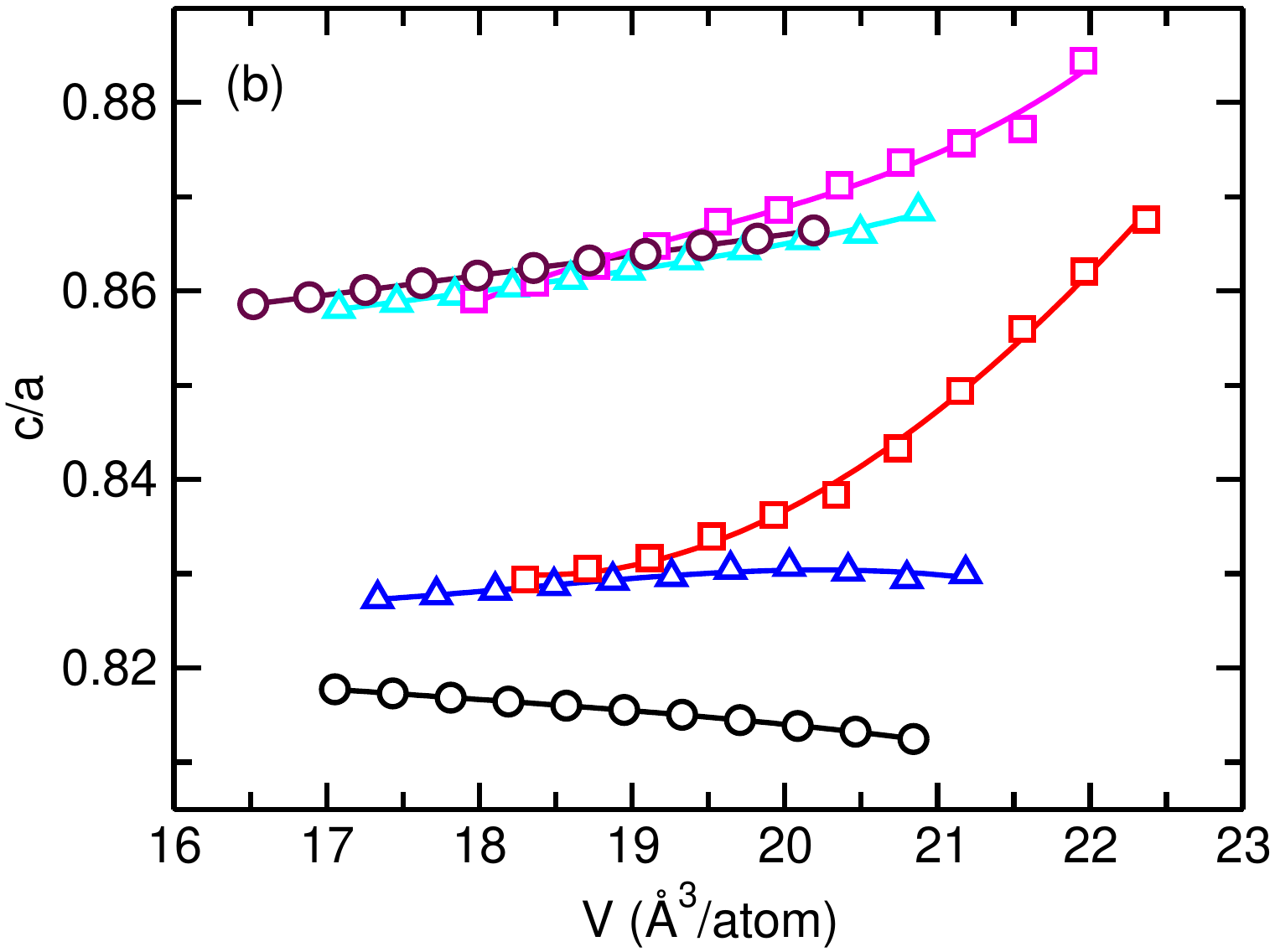}~
\caption{(a) Total energy difference per atom $\Delta E$ and (b) optimized lattice constants ratio $c/a$ as a function of cell volume per atom in tetragonal intermetallics \ce{FeSn2}, \ce{FeLiSn2}, \ce{FeNaSn2}, \ce{CoSn2}, \ce{CoLiSn2} and \ce{CoNaSn2}.}
\label{fig2}
\end{figure}
To determine the optimized lattice parameters of the unit cells with these magnetic orderings, we plotted the energy-volume curves by calculating the total energies as gradually increasing the unit cell volume.
Here, the optimized tetragonal ratio $c/a$ at each volume was determined by calculating total energy as varying $c/a$ with allowing the relaxation of atomic positions and by interpolating the data to the third-order polynomial (see Fig. S1).
The obtained energy-volume curves for all the intermetallics studied in this work are shown in Fig.~\ref{fig2}(a) and the optimized tetragonal ratios as increasing the cell volume are plotted in Fig.~\ref{fig2}(b).
Table~\ref{tabl1} lists the determined lattice parameters and mass density of these intermetallics.
For the cases of \ce{FeSn2} and \ce{CoSn2}, the calculated lattice constants were found to be agreed well with the experimental values~\cite{Armbruster07zkn} with very low values of relative error under 1\%.

It was found that replacing one M atom in the \ce{MSn2} unit cell with A atom induces an increase of lattice constants, tetragonal ratio and thus unit cell volume and decrease of density.
The volume expansion rates $V/V_0\times100$\% ($V_0$ the volume of \ce{MSn2} unit cell) for \ce{FeASn2} were estimated to be smaller (104.5\% for Li and 111.4\% for Na) than those for \ce{CoASn2} (106.6 and 112.6\%).
Accordingly, the relative density decrease rates $(\rho-\rho_0)/\rho_o\times100$\% for \ce{FeASn2} ($-8.3$\% for Li and $-12.8$\% for Na) were found to be smaller than those for \ce{CoASn2} ($-10.3$ and $-13.8$\%).
Such volume expansion is mainly associated with an increase of interatomic distance between the M atoms from 2.663/2.736 \AA~in \ce{MSn2} to 2.736/2.798 \AA~for Li and to 2.837/2.869 \AA~for Na exchanges respectively, indicating a weakening of interatomic reaction by alkali metal exchange.
Note that the M$-$Sn distance also increases but the Sn$-$Sn distance slightly decreases by such exchange (see Table S3).
The volume expansion and relative interatomic distance indicate that the bonding strength of Fe$-$Fe is stronger than that of Co$-$Co, and replacing Fe or Co with Li atom makes the bonding strength weaker than with Na atom.

To gain insight into the structural stability, we evaluated the cohesive energy $E_c=E_{\text{tot}}-\sum_iE_{i}^{\text{atom}}$ and the formation energy $E_f=E_{\text{tot}}-\sum_iE_{i}^{\text{bulk}}$, where $E_{i}^{\text{atom}}$ and $E_{i}^{\text{bulk}}$ are the total energies of isolated atom and elementary bulk of $i$-th species.
As shown in Table~\ref{tabl1}, the calculated cohesive and formation energies are negative for all the compounds, indicating that these phases are thermodynamically stable and can be formed at ambient condition from their elemental constituents.
The formation energy of \ce{FeSn2} ($-0.221$ eV/atom) is lower than that of \ce{CoSn2} ($-0.183$ eV/atom), demonstrating that the Fe-related intermetallics is more stable than the Co-related ones, which is consistent with the bonding strength analysis though being contrast to the previous DFT calculation~\cite{Sun16cms}.
It was also observed that substituting Li or Na for Fe or Co atom makes the compounds less stable, and this effect is more pronounced for Na substitution compared to Li substitution.

\subsection{Elastic and mechanical properties}
\begin{table*}[!th]
\small
\caption{Elastic stiffness constants $C_{ij}$, bulk modulus $B$, shear modulus $G$, Young's modulus $E$, Pugh's ratio $B/G$, Poisson's ratio $\nu$, average sound velocity $\upsilon_m$, and Debye temperature $\theta_{\text{D}}$ of the considering intermetallics.}
\label{tabl2}
\begin{tabular}{lcccccc@{}cccccccc}
\hline
& \multicolumn{6}{c}{Elastic stiffness constant (GPa)} & & \multicolumn{3}{c}{Elastic modulus (GPa)} & & & & \\
\cline{2-7} \cline{9-11}
Compound & $C_{11}$ & $C_{12}$ & $C_{13}$ & $C_{33}$ & $C_{44}$ & $C_{66}$ & & $B$ & $G$ & $E$ & $B/G$ & $\nu$ & $\upsilon_m$ (m/s) & $\theta_{\text{D}}$ (K) \\
\hline
\ce{FeSn2} & 202.4 & 48.6 & 73.5 & 178.5 & 53.4 & 69.2 & & 108.3 & 60.3 & 152.6 & 1.796 & 0.265 & 2950.6 & 706.8 \\
\ce{FeSn2}$^a$ & 184.9 & 22.8 & 43.7 & 190.6 & 42.8 & 62.7 & & 86.6 & 57.7 & 142.7 & 1.501 & & & \\
\ce{Fe_{3/4}Li_{1/4}Sn2} & 164.1 & 41.4 & 45.7 & 141.7 & 45.2 & 54.9 & & 81.6 & 51.2 & 127.0 & 1.595 & 0.241 & 2830.6 & 668.2 \\
\ce{Fe_{3/4}Na_{1/4}Sn2} & 146.1 & 43.1 & 40.9 & 123.9 & 42.6 & 51.0 & & 73.7 & 46.5 & 115.2 & 1.587 & 0.240 & 2765.2 & 638.9 \\
\hline
\ce{CoSn2} & 204.1 & 52.9 & 58.3 & 190.1 & 42.7 & 67.0 & & 104.1 & 57.4 & 145.4 & 1.815 & 0.267 & 2818.3 & 682.3 \\
\ce{CoSn2}$^a$ & 208.5 & 53.8 & 62.4 & 178.7 & 43.2 & 68.3 & & 105.8 & 57.1 & 145.2 & 1.853 & & & \\
\ce{Co_{3/4}Li_{1/4}Sn2} & 151.0 & 37.8 & 44.3 & 130.5 & 28.1 & 46.0 & & 76.1 & 39.3 & 100.5 & 1.937 & 0.280 & 2466.7 & 584.6 \\
\ce{Co_{3/4}Na_{1/4}Sn2} & 136.7 & 36.9 & 41.8 & 107.7 & 25.6 & 39.7 & & 68.8 & 34.4 & 88.4 & 2.002 & 0.286 & 2356.5 & 548.4 \\
\hline
\end{tabular} \\
$^a$ DFT calculation data with PBE functional~\cite{Sun16cms}.
\end{table*}
The mechanical stability of intermetallics can be estimated by its elastic constants, which play an important role in describing the resistance of material against mechanical deformation.
The calculated six independent elastic stiffness constants for \ce{MSn2} and \ce{MASn2} in the tetragonal phase, namely $C_{11}$, $C_{12}$, $C_{13}$, $C_{33}$, $C_{44}$ and $C_{66}$, are listed in Table~\ref{tabl2}.
For the cases of \ce{MSn2}, our calculated data is in overall agreement with the previous first-principles calculation~\cite{Sun16cms}.
These independent elastic constants were found to satisfy the well-known Born stability criteria for tetragonal crystal~\cite{Max56}, expressed by $C_{11}>0$, $C_{33}>0$, $C_{44}>0$, $C_{66}>0$, $C_{11}-C_{12}>0$, $C_{11}+C_{33}-2C_{13}>0$, $2(C_{11}+C_{12})+C_{33}+4C_{13}>0$, thereby implying their mechanical stability at zero pressure.
For all these alloys, $C_{11}$ is larger than $C_{33}$, indicating that their deformation resistance along the $a$-axis is stronger than that along the $c$-axis.
In accordance with the above discussion, substituting Li or Na atom was confirmed to reduce the mechanical stability and deformation resistance along both the $a$- and $c$-axises, due to lowering of elastic constants, and this effect was more influential in Na replacement than Li one.
It should be noted that the shear deformation resistance is weakened as well by such substitution due to smaller values of $C_{44}$ in \ce{MASn2} than in \ce{MSn2}.

As the strength of a polycrystalline solid is estimated by its elastic modulus, which can be readily evaluated from the elastic constants (see Table S4 for elastic compliance constants), we present the bulk, shear and Young's moduli in Table~\ref{tabl2}.
The bulk and shear moduli within the Voigt ($B_{\text{V}}$, $G_{\text{V}}$) and Reuss ($B_{\text{R}}$, $G_{\text{R}}$) approximations are shown in Table S5.
Since the calculated elastic moduli are much lower than 360 GPa, these alloys are not said to be incompressible~\cite{Chiodo06cpl}.
It was known that the bulk, shear and Young's moduli describe the material response to uniform pressure, shear and uniaxial stress respectively.
In accordance with this fact, Table~\ref{tabl2} shows that the elastic moduli of Fe-related intermetallics are larger than those of Co-related ones, indicating that the former has stronger resistance to volume compression and more prominent directional bonding between the constituent atoms than the latter.

According to the Pugh criteria for ductility of solid~\cite{Pugh54pm}, the critical values of Pugh's ratio $B/G$ and Poisson's ratio $\nu$ are 1.75 and 0.26; when the $B/G$ value is greater than 1.75 or the $\nu$ is larger than 0.26, the crystal is considered as a ductile material, otherwise it is a brittle material~\cite{Hadi17cms}.
It was found that \ce{FeSn2} and \ce{CoSn2} are ductile materials due to their $B/G$ values of 1.796 and 1.815 being larger than 1.75 and $\nu$ values of 0.265 and 0.267 being larger than 0.26.
Interestingly, Li or Na substitution for Fe in \ce{FeSn2} induces decreases of $B/G$ and $\nu$ values to 1.595 or 1.587 and 0.241 or 0.240 respectively, thereby indicating a transition from ductile to brittle property.
However, an opposite trend is observed in \ce{CoSn2}; Li or Na substitution increases the $B/G$ and $\nu$ values to 1.937 or 2.002 and 0.280 or 0.286 respectively, suggesting that Li or Na substitution increases the ductility.

In Table~\ref{tabl2}, we also present the average sound velocity and Debye temperature (see Table S5 for longitudinal and transverse elastic wave velocities).
The Debye temperature $\theta_{\text{D}}$ is known to be associated with the vibration of atoms and hardness of a solid; the higher $\theta_{\text{D}}$ value implies the stronger interaction between atoms and the higher hardness.
Our calculation result shows that \ce{FeSn2} (706.8 K) has higher Debye temperature than \ce{CoSn2} (682.3 K), and moreover, Li or Na substitution reduces the $\theta_{\text{D}}$ values.
Such tendency indicates that \ce{FeSn2} is harder than \ce{CoSn2} and Li or Na substitution lowers the hardness, which is consistent with the variation of elastic modulus.

\subsection{Lattice vibrational properties}
\begin{figure}[!b]
\centering
\includegraphics[clip=true,scale=0.5]{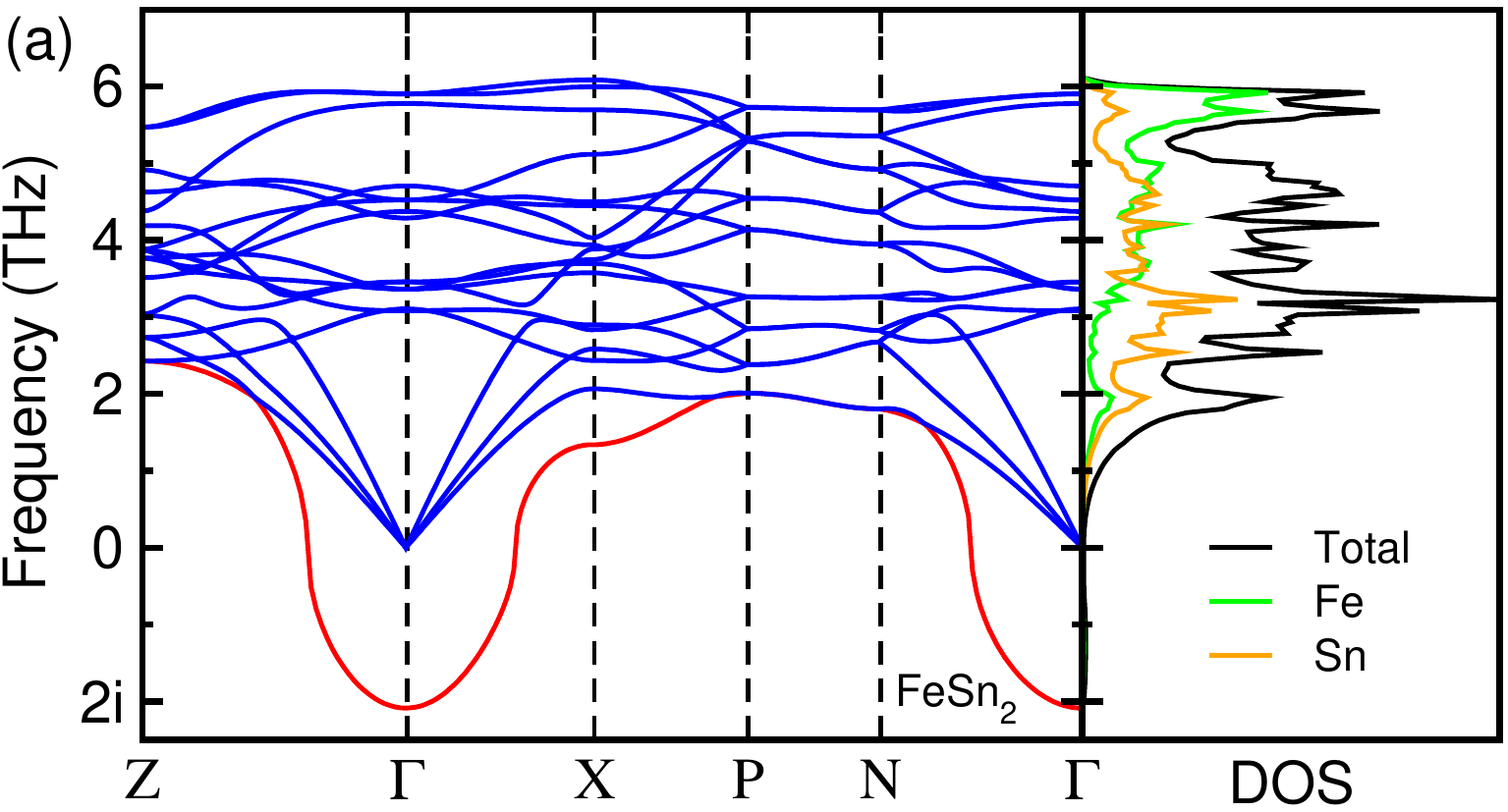}
\includegraphics[clip=true,scale=0.5]{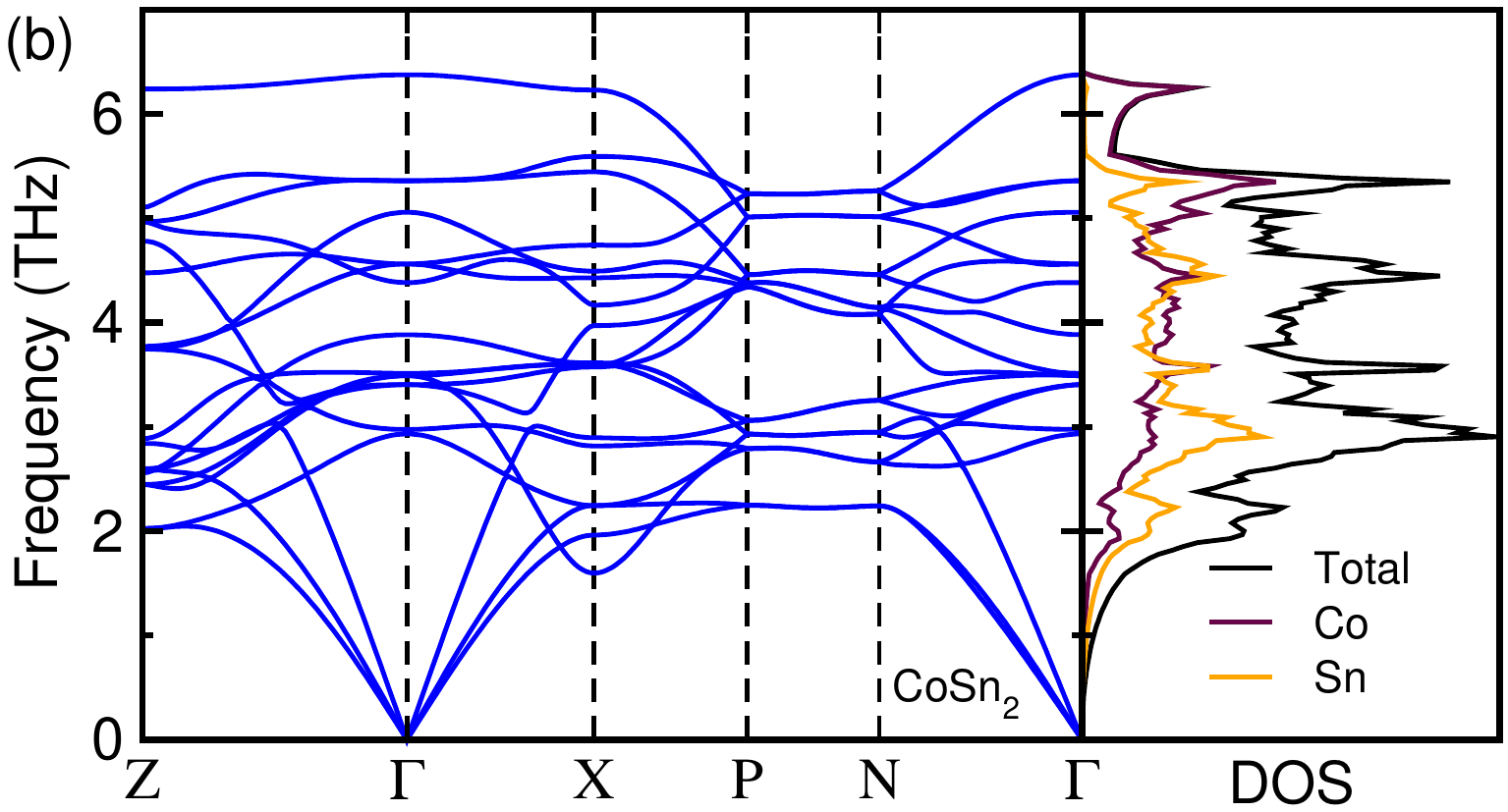}
\caption{Phonon dispersion curves and the atomic resolved phonon density of states for (a) \ce{FeSn2}, which has soft mode with imaginary phonon frequency indicated by red color, and (b) \ce{CoSn2}. The primitive unit cell including two formula units was used in the calculation.}
\label{fig3}
\end{figure}
\begin{figure*}[!th]
\centering
\includegraphics[clip=true,scale=0.5]{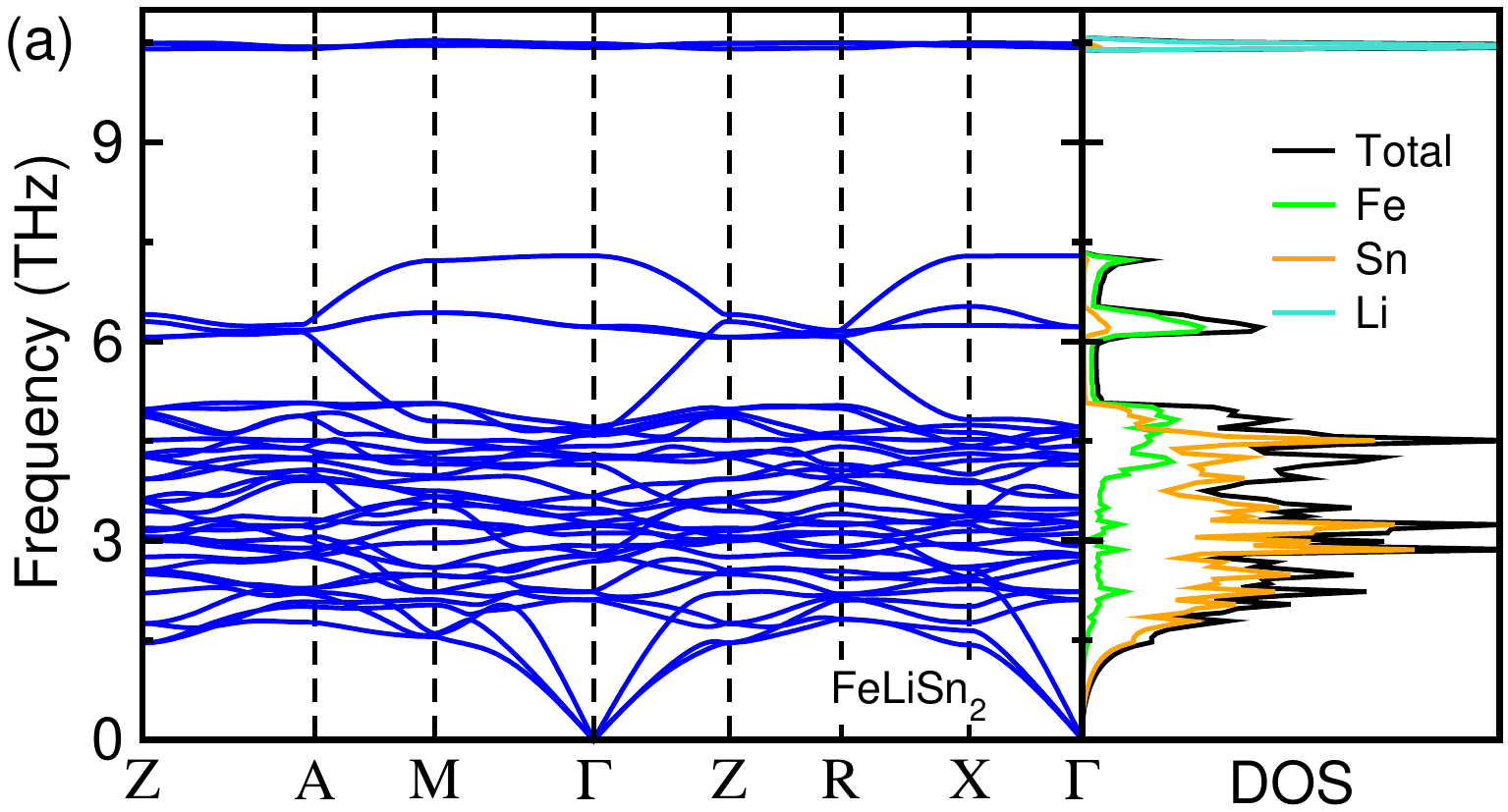}~~
\includegraphics[clip=true,scale=0.5]{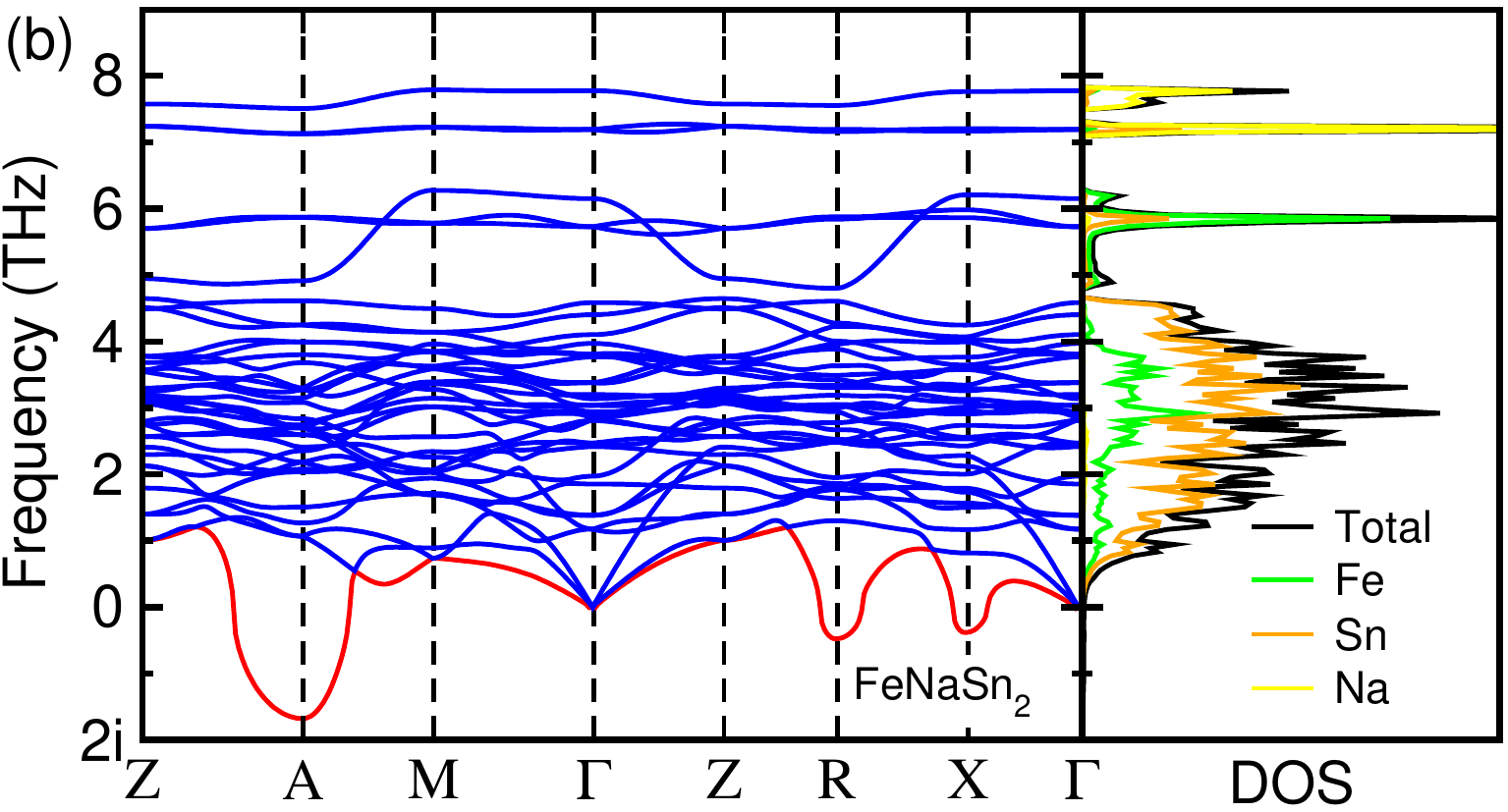} \\
\includegraphics[clip=true,scale=0.5]{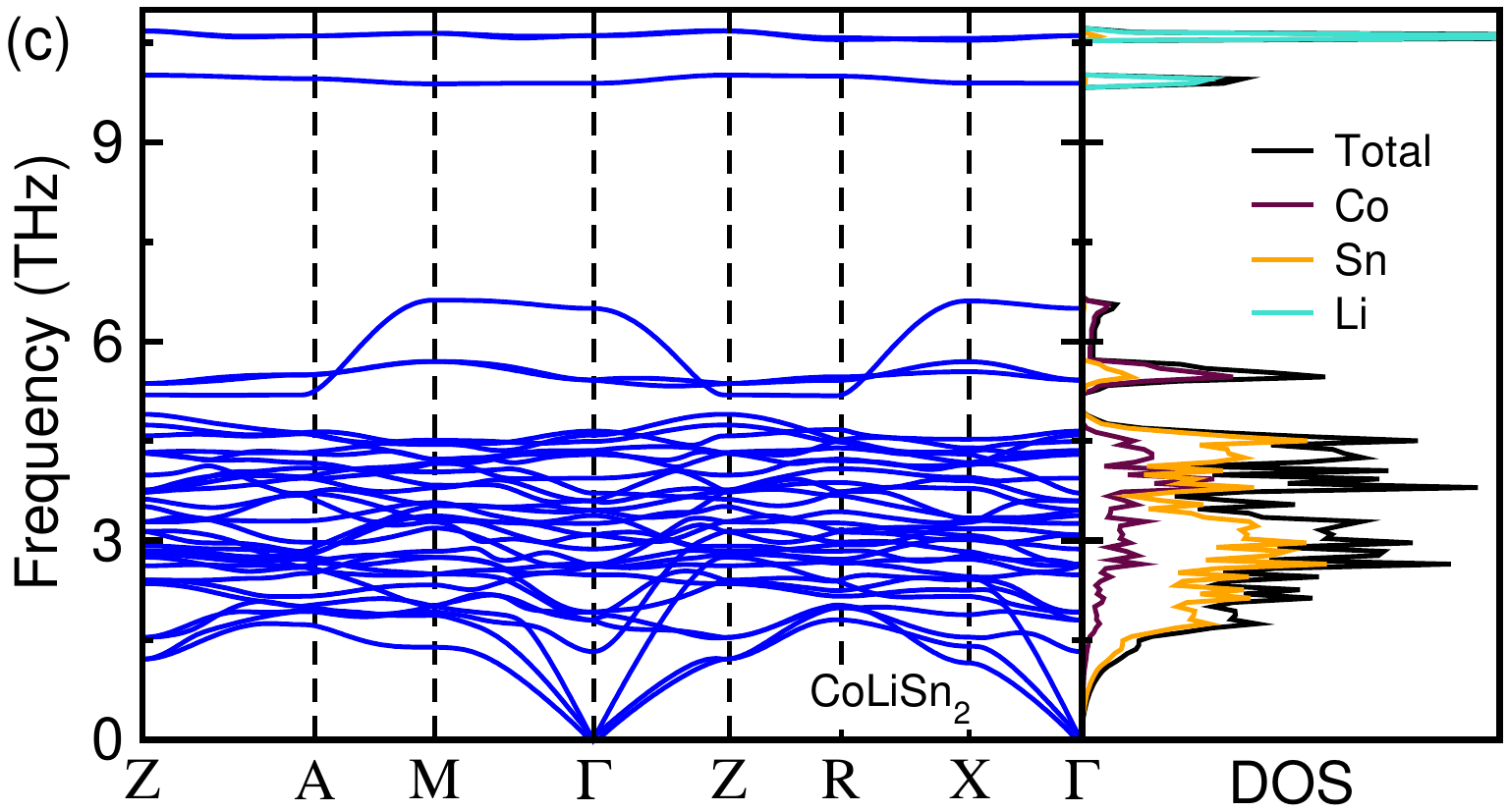}~~
\includegraphics[clip=true,scale=0.5]{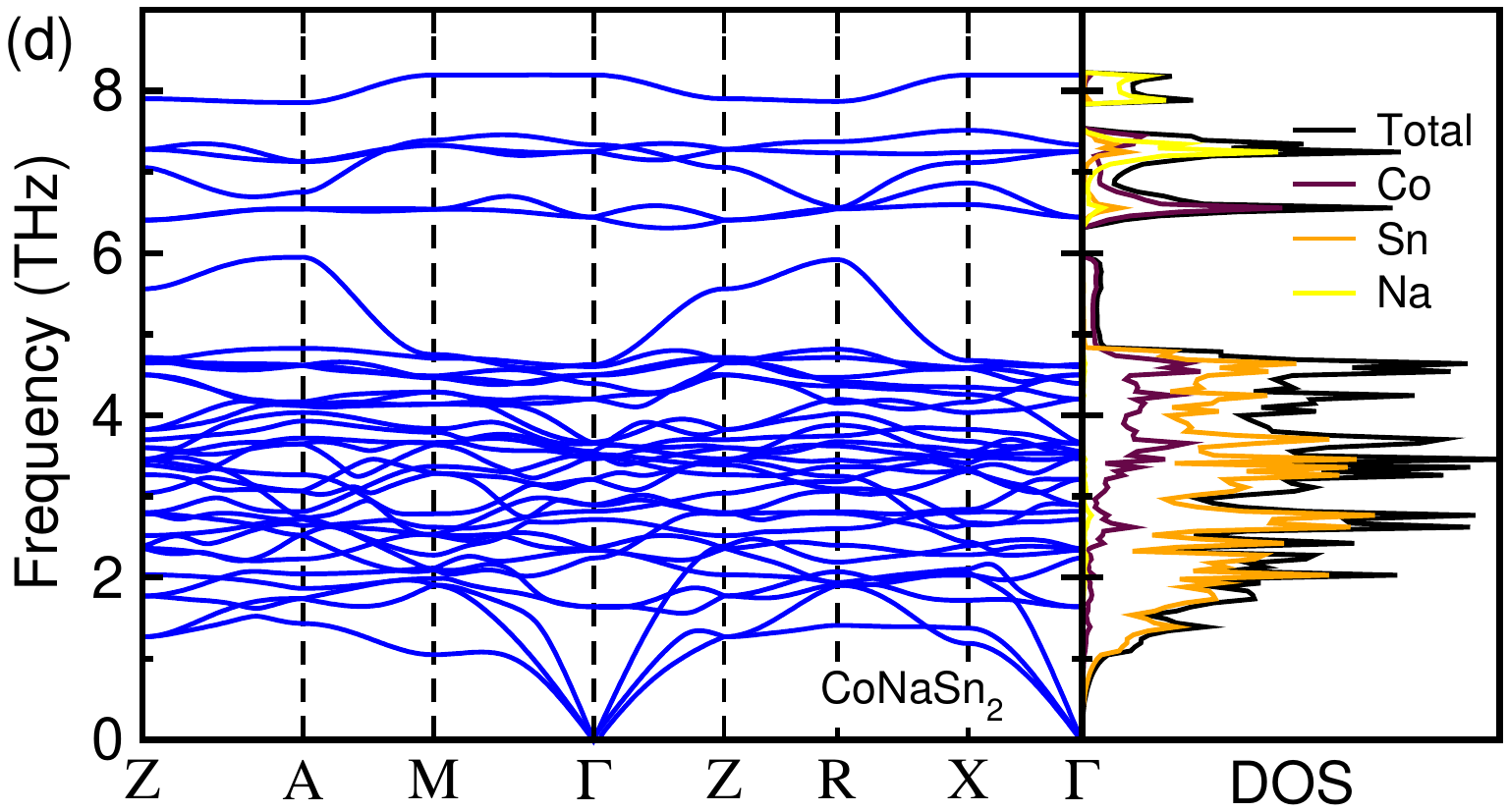}
\caption{Phonon dispersion curves and the atomic resolved phonon DOS for (a) \ce{FeLiSn2}, (b) \ce{FeNaSn2}, (c) \ce{CoLiSn2} and (d) \ce{CoNaSn2}. For the case of \ce{FeNaSn2}, an imaginary phonon mode originated from acoustic phonon appears around A, R and X points, as indicated by red line.}
\label{fig4}
\end{figure*}

To further discuss the dynamic stability of the intermetallic compounds, the phonon dispersion curves and the corresponding phonon DOS were plotted.
Figure~\ref{fig3} depicts those for \ce{FeSn2} and \ce{CoSn2}.
It turned out that there is no imaginary phonon mode in \ce{CoSn2} but a non-degenerate optical phonon mode with imaginary phonon frequency values appears around the zone-center $\Gamma$ point for the case of \ce{FeSn2}, suggesting that at ambient condition \ce{CoSn2} is dynamically stable but \ce{FeSn2} should transformed to another phase.
From the analysis of phonon DOS for \ce{FeSn2}, the imaginary soft phonon was found to be mainly caused by the relative vibration among Fe atoms.
When allowing atom movement along the soft phonon vector, the Fe atoms moved along the $c$-axis direction, resulting in an extension of Fe$-$Fe interatomic distance and furthermore phase transition from tetragonal $I4/mcm$ to $I4/m$ space group phases.
Such space group change has been already illustrated for \ce{Mo2B} with first-principles calculations by Zhou and co-workers~\cite{Zhou14jap}.

Then the influence of M/A substitution on lattice vibrational properties of \ce{FeSn2} and \ce{CoSn2} was explored along the above discussion.
Figure~\ref{fig4} shows the phonon dispersion curves with phonon DOS for \ce{MASn2} in tetragonal $P422$ space group phase, containing 12 atoms.
One interesting finding is that for \ce{FeNaSn2} an imaginary phonon mode originated from acoustic phonon appears around A, R and X points, indicating that this phase is dynamically unstable like \ce{FeSn2}.
However, other intermetallic compounds were found to be stable at ambient condition due to no imaginary phonon mode.
For each \ce{MASn2}, there is a gap in the region between low frequency and high frequency.
The gaps in Li-substituted compounds are larger than those in Na-substituted ones.
The high frequency region above 6 THz is predominantly contributed by Li or Na atoms, while the contribution is derived from vibrations of Fe or Co and Sn atoms below 6 THz.

\subsection{Electronic properties}

\begin{figure}[!b]
\centering
\begin{tabular}{c@{\hspace{0pt}}c}
\includegraphics[clip=true,scale=0.5]{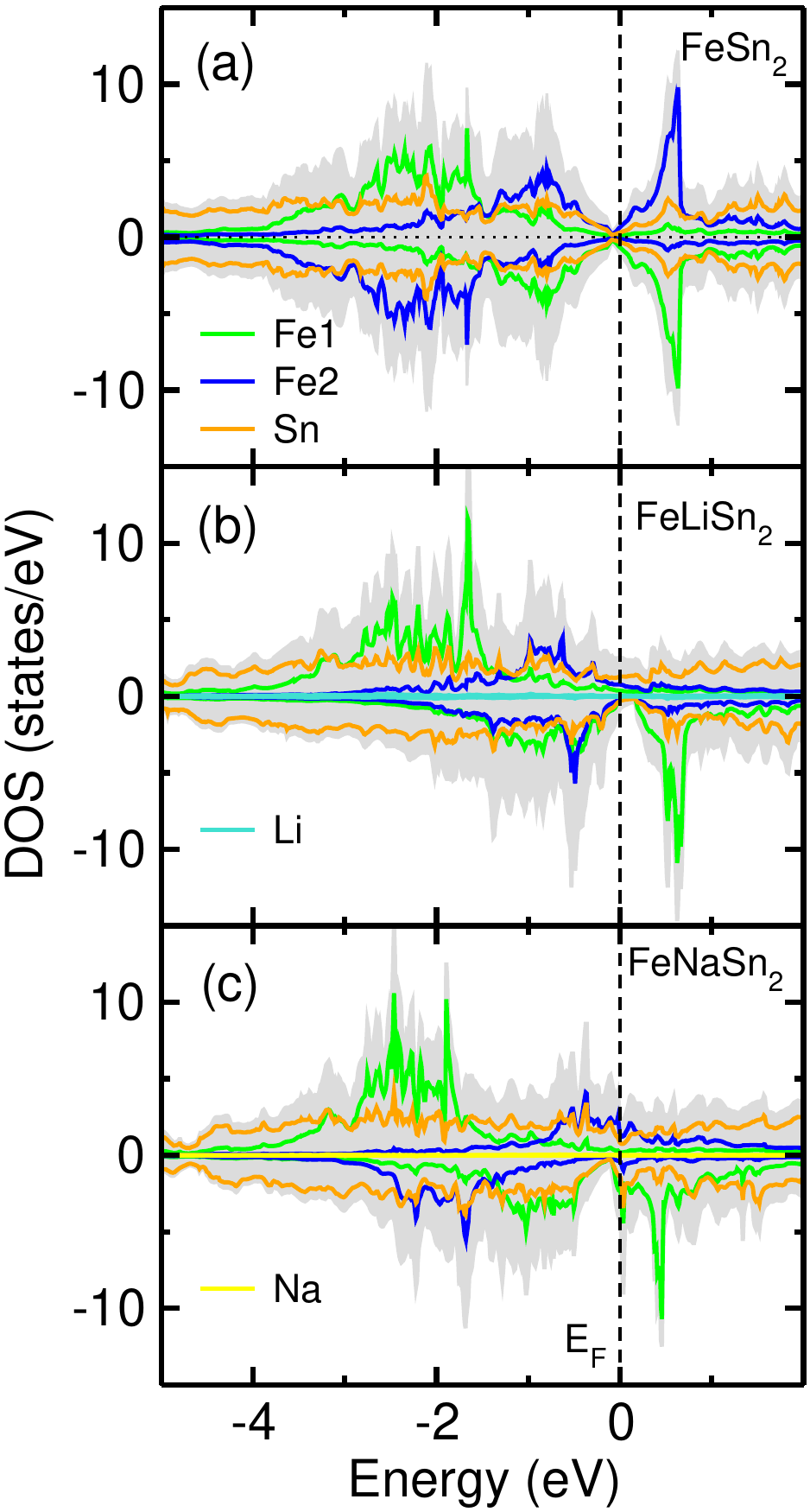} &
\includegraphics[clip=true,scale=0.5]{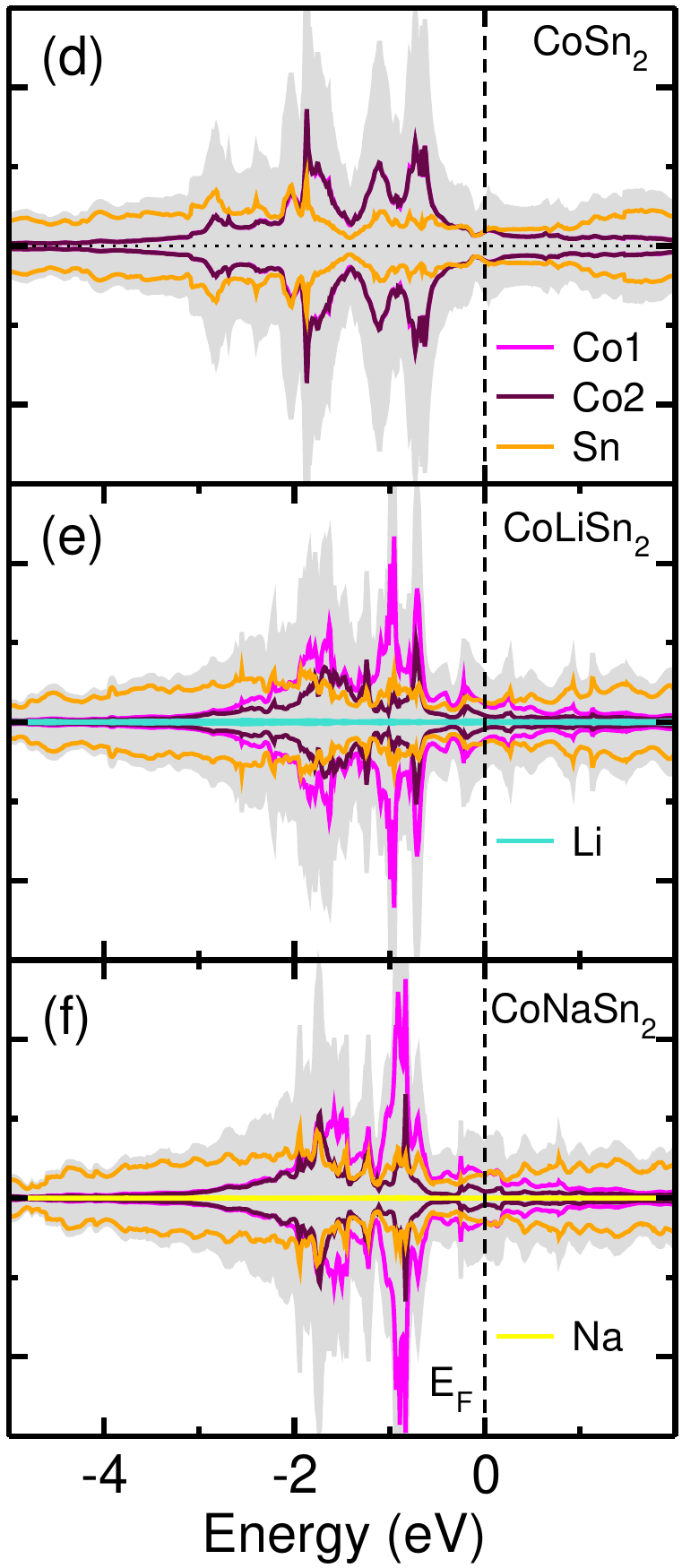}
\end{tabular}
\caption{Atomic resolved density of states (DOS) for (a) \ce{FeSn2}, (b) \ce{FeLiSn2}, (c) \ce{FeNaSn2}, (d) \ce{CoSn2}, (e) \ce{CoLiSn2} and (f) \ce{CoNaSn2}. Gray-colored filled region denotes the total DOS, and the Fermi level $E_{\text{F}}$ is set zero, as indicated by vertical dashed line.}
\label{fig5}
\end{figure}
\begin{figure*}[!th]
\centering
\includegraphics[clip=true,scale=0.5]{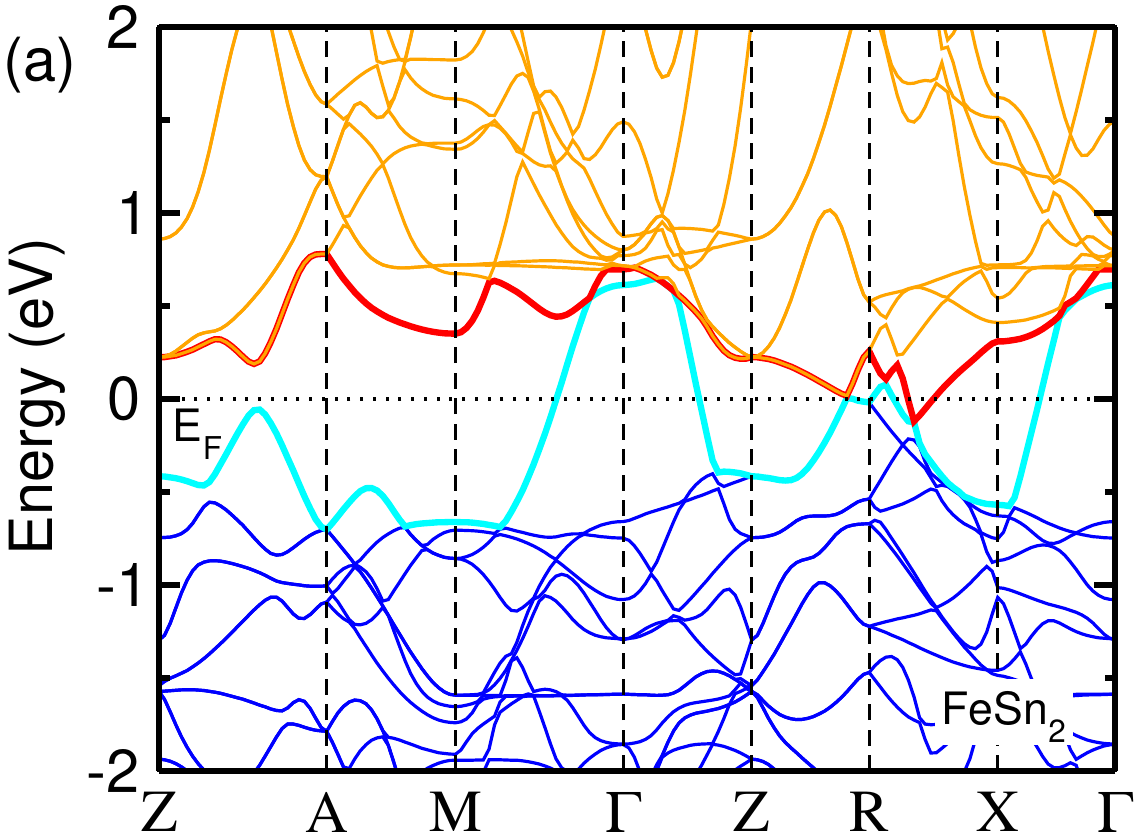}
\includegraphics[clip=true,scale=0.5]{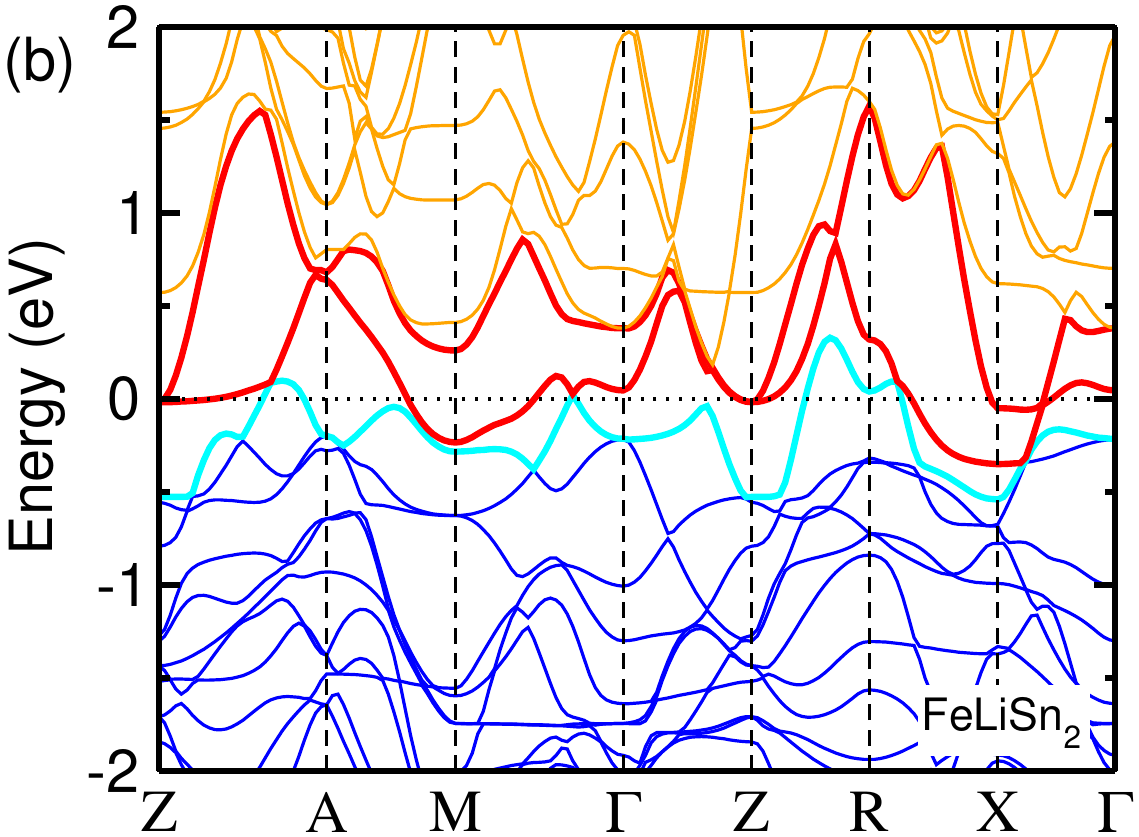}
\includegraphics[clip=true,scale=0.5]{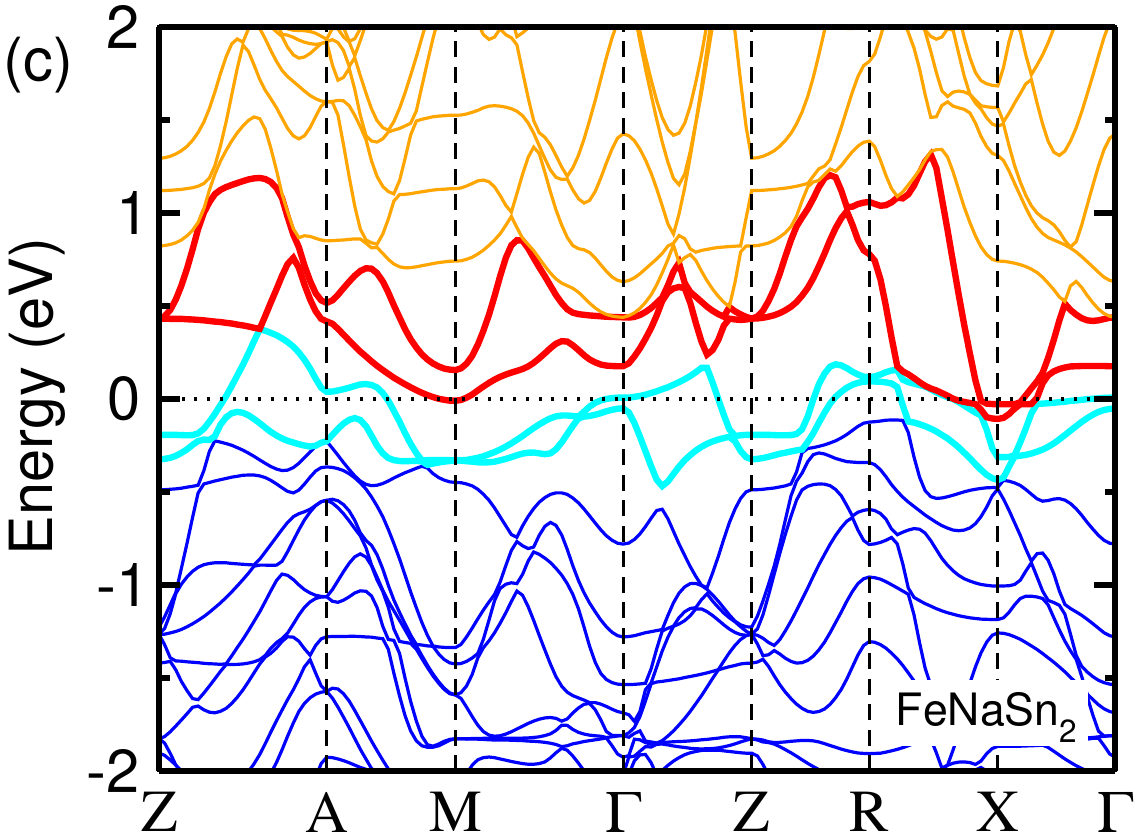} \\
\includegraphics[clip=true,scale=0.5]{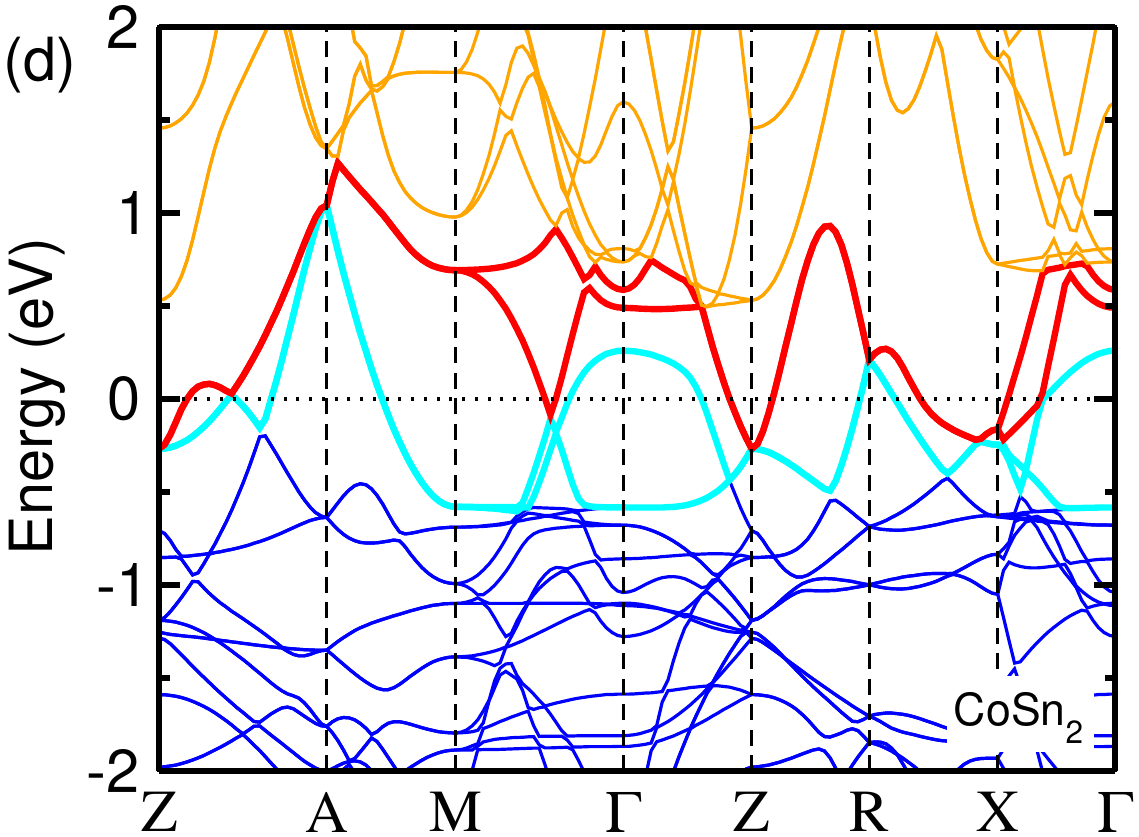}
\includegraphics[clip=true,scale=0.5]{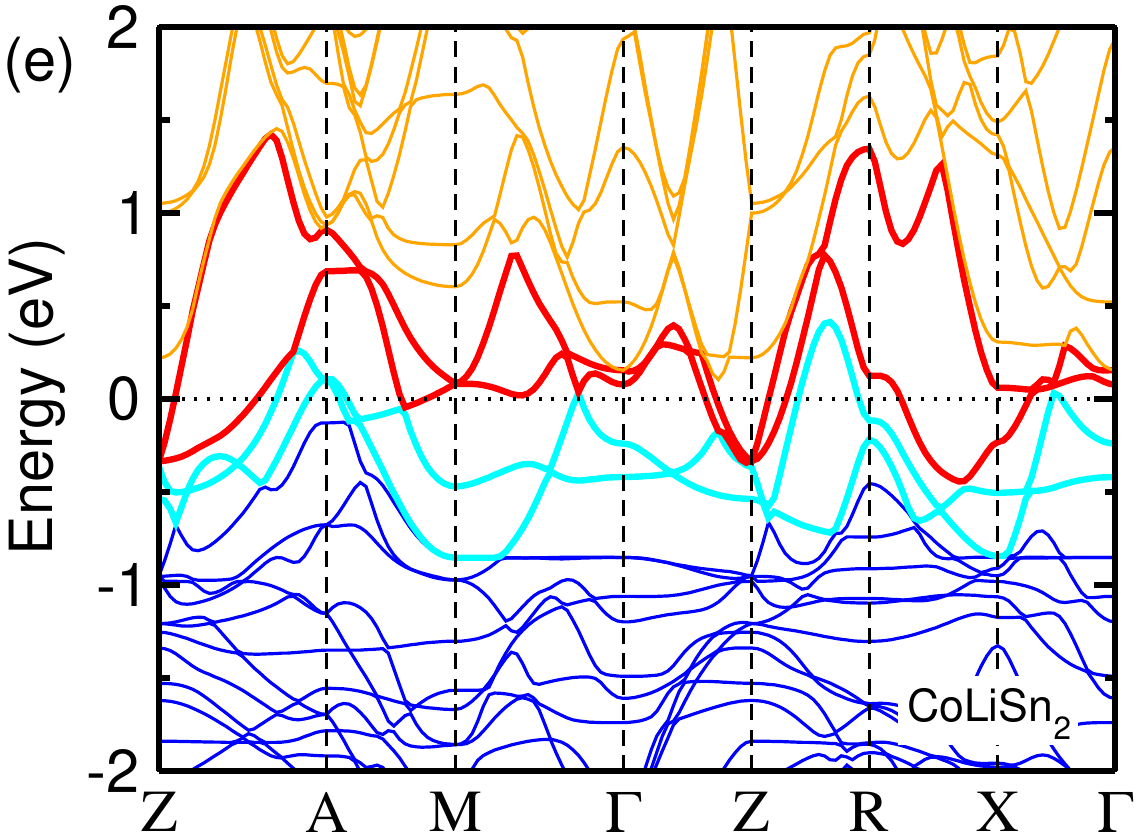}
\includegraphics[clip=true,scale=0.5]{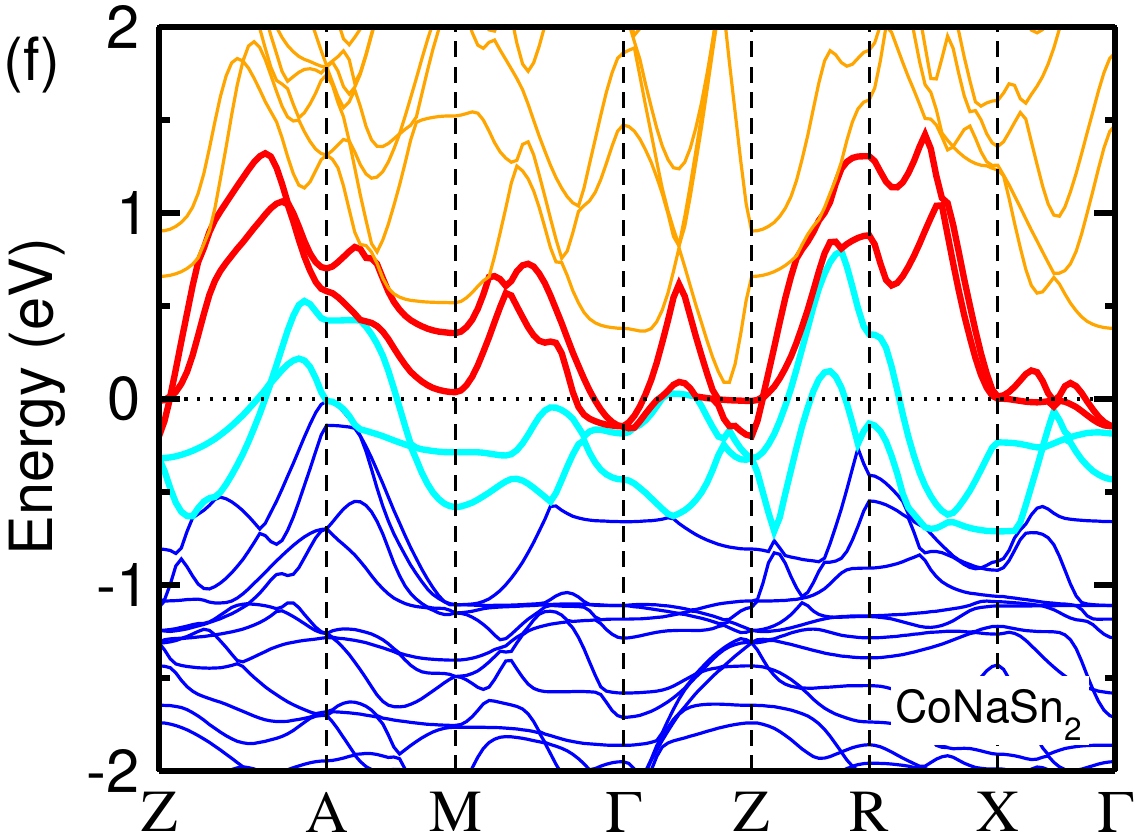}
\caption{Energy band structures in (a) \ce{FeSn2}, (b) \ce{Fe_{3/4}Li_{1/4}Sn2}, (c) \ce{Fe_{3/4}Na_{1/4}Sn2}, (d) \ce{CoSn2}, (e) \ce{Co_{3/4}Li_{1/4}Sn2} and (f) \ce{Co_{3/4}Na_{1/4}Sn2}, calculated by using the PBE exchange-correlation functional. Spin polarization is considered and only spin-up bands are shown here. Blue- and orange-colored lines indicate the occupied and empty bands, while cyan- and red-colored thick lines denote the bands crossing the Fermi level $E_{\text{F}}$, which is set to zero as indicated by horizontal dotted line.}
\label{fig6}
\end{figure*}

\begin{figure}[!t]
\centering
\includegraphics[clip=true,scale=0.09]{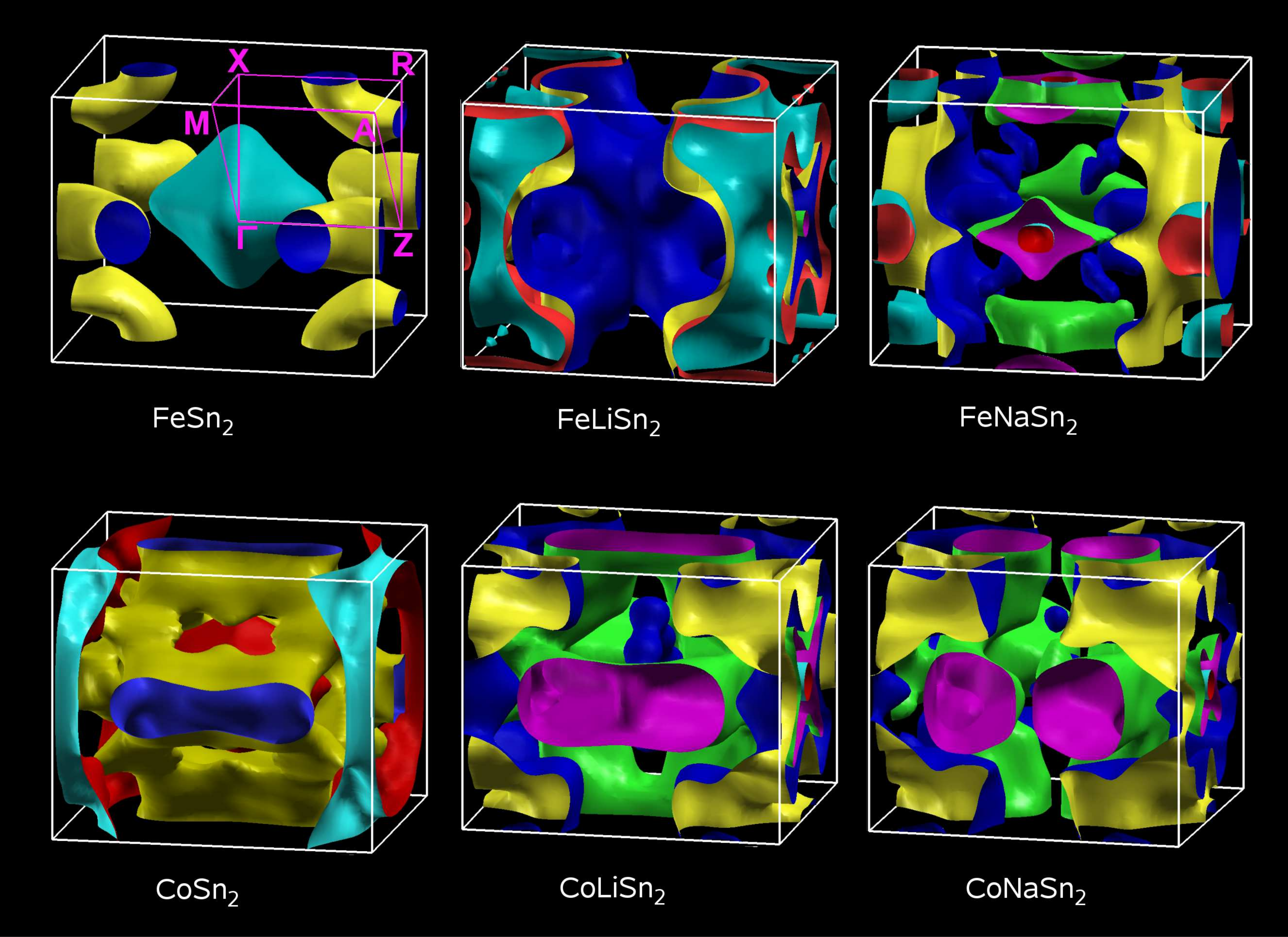}
\caption{Fermi surface for \ce{MSn2} (M = Fe, Co) and \ce{MASn2} (A = Li, Na). Pink-colored labels indicate the high-symmetry point of first Brillouin zone.}
\label{fig7}
\end{figure}

In order to understand the electronic properties, we calculated the electronic density of states (DOS) and energy band structures of \ce{MSn2} and \ce{MASn2} with spin-polarization.
Figure~\ref{fig5} shows the atomic resolved total DOS around the Fermi level $E_{\text{F}}$, which is set to zero as indicated by vertical dashed line, in the energy range from $-5$ eV to 2 eV (see Fig. S2 for orbital resolved partial DOS).
We show the spin-up band structures in the energy range of interest ($-2$ eV to 2 eV) along the high-symmetry lines of the first Brillouin zone in Fig.~\ref{fig7}.
It should be noted that for the compounds with spin-ordering of AFM and NM, which are \ce{FeSn2}, \ce{CoSn2}, \ce{CoLiSn2} and \ce{CoNaSn2}, the spin-down bands perfectly coincide with the spin-up bands, while for \ce{FeLiSn2} and \ce{FeNaSn2} in the FM state the spin-down bands are strikingly different from the spin-up bands (see Fig. S3 for comparison between spin-up and spin-down bands in \ce{FeLiSn2} and \ce{FeNaSn2}).

For all the compounds, we can see 2$\sim$4 bands crossing the Fermi level, denoted by red- and cyan-colored thick lines in Fig.~\ref{fig6}, accounting for the metallic behavior.
The number of crossing bands and atomic contribution characteristics vary with the compound.
For the case of \ce{FeSn2}, one hole (cyan color) and one electron (red color) bands appear across the Fermi level, of which the hole band is contributed from Fe-$3d$ and Sn-$5s, 5p$ states almost equally but the electron band above $E_{\text{F}}$ is dominated by Fe-$3d$ state, as shown in Fig.~\ref{fig5}(a) and Fig. S2(a).
Contrastingly, \ce{CoSn2} exhibits two hole and two electron bands across the Fermi level, of which the hole bands below $E_{\text{F}}$ are dominated by Co-$3d$ with a slight contribution from Sn-$5p$ states but the electron bands are contributed equally from Co-$3d$ and Sn-$5p$ states, as shown in Fig.~\ref{fig5}(d) and Fig. S2(d).
For M/A exchange, \ce{FeLiSn2} has 3 crossing (two electron and one hole) bands (Fig.~\ref{fig6}(b)), and others also have 4 (two hole and two electron) bands crossing the Fermi level (Fig.~\ref{fig6}(c), (e), (f)).
Similar contribution characteristics to \ce{MSn2} is observed for \ce{MASn2}.
However, since one M2 atom with minus magnetization was replaced by alkali atom, enhanced contribution from M1 atom with plus magnetization compared to M2 atom can be seen in Fig.~\ref{fig5} for total DOS.
It is worth noting that from the total DOS plot, \ce{FeSn2} is clearly in AFM state and \ce{FeASn2} is in FM state, whereas \ce{CoSn2} and \ce{CoASn2} have NM feature.

These band structure features yield a multi-sheet Fermi surface, as shown in Fig.~\ref{fig7}.
There are several kinds of shape for the Fermi surface, including round corner polyhedron, rugged ring, quasi-2D sheet, and wrinkled and branched cylinder.
The relatively simple topology of Fermi surface, Fermi hole pocket with a shape of round polyhedron at zone center around $\Gamma$ point and bent cylinder for electron at zone boundary going from Z to R, is observed for \ce{FeSn2}, while complicated shapes appear for \ce{CoSn2}.
For the case of \ce{FeASn2}, we can see wavy-like 2D sheet for holes and electron pockets at the center of zone boundary surface, while for \ce{CoASn2} the electron Fermi surface with a shape of wrinkled cylinder is characteristically observed.

\subsection{Defect formation and alkali atom migration}
In order to simulate the initial interaction of alkali metal with \ce{MSn2}, we considered the point defect such as vacancies of $V_{\ce{M}}$ and $V_{\ce{Sn}}$ and  substitutional solutes of $\ce{A}_{\ce{M}}$ and $\ce{A}_{\ce{Sn}}$.
The defect formation energy was calculated using Eq.~\ref{eq_df} and the formation enthalpy of the compound with such defect was calculated using Eq.~\ref{eq_hf}.
Table~\ref{tabl3} presents the calculated energetic values for these defects.
The formation enthalpies for all the kinds of defect were calculated to be negative for both \ce{FeSn2} and \ce{CoSn2}, indicating that the compounds with defect can be formed in exothermic way.
The reactivity of \ce{CoSn2} with Li or Na was found to be higher than that of \ce{FeSn2} due to smaller values of formation enthalpy.
It was found that for both \ce{FeSn2} and \ce{CoSn2}, the formation energy of Sn vacancy is lower than those of Fe or Co vacancies, and Li substitutional solute has lower formation energy than Na ones.
In particular, for the case of \ce{CoSn2}, Sn-related defects such as $V_{\ce{Sn}}$,  $\ce{Li}_{\ce{Sn}}$ and  $\ce{Na}_{\ce{Sn}}$ have negative formation energies of $-0.662$, $-2.307$ and $-1.539$ eV respectively.
These suggest that Sn atom has higher reactivity with Li or Na than Fe or Co atom and moreover lithium is more reactive with \ce{MSn2} than sodium.
\begin{table}[!t]
\small
\caption{Defet formation energy ($E_f$) and formation entalpy per atom ($\Delta H$) of compound with defect, including vacancies ($V_{\ce{M}}$, $V_{\ce{Sn}}$) and substitutional solutes ($\ce{A}_{\ce{M}}$, $\ce{A}_{\ce{Sn}}$) in intermetallic compounds \ce{MSn2} (M = Fe, Co; A = Li, Na).}
\label{tabl3}
\begin{tabular}{lcc@{}ccc}
\hline
 & \multicolumn{2}{c}{\ce{FeSn2}} & & \multicolumn{2}{c}{\ce{CoSn2}} \\
\cline{2-3} \cline{5-6}
Defect & $E_f$ (eV) & $\Delta H$ (eV) & & $E_f$ (eV) & $\Delta H$ (eV) \\
\hline
$V_{\ce{M}}$  & 2.876 & $-1.487$ & & ~~~4.590 & $-2.136$ \\
$V_{\ce{Sn}}$          & 2.193 & $-1.494$ & & $-0.662$ & $-2.191$ \\
Li$_{\ce{M}}$ & 2.146 & $-1.479$ & & ~~~4.067 & $-2.119$ \\
Na$_{\ce{M}}$ & 3.998 & $-1.460$ & & ~~~5.643 & $-2.103$ \\
Li$_{\ce{Sn}}$         & 0.784 & $-1.493$ & & $-2.307$ & $-2.186$ \\
Na$_{\ce{Sn}}$         & 1.511 & $-1.486$ & & $-1.539$ & $-2.178$ \\
\hline
\end{tabular}
\end{table}
\begin{figure}[!b]
\centering
\includegraphics[clip=true,scale=0.11]{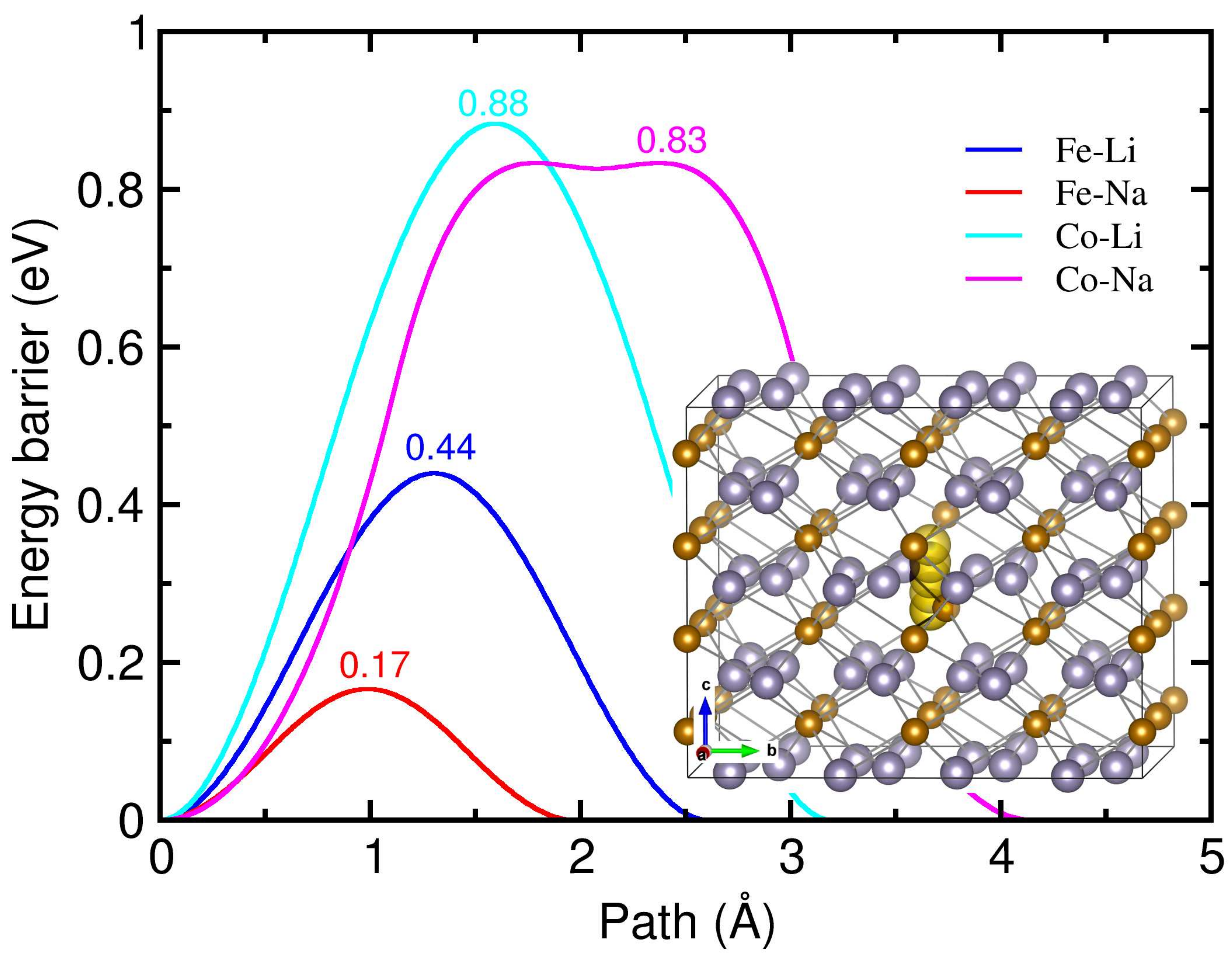}
\caption{Activation energy barrier for vacancy-mediated Li/Na migration in \ce{FeSn2} and \ce{CoSn2}. Inset shows the migration path denoted by yellow-colored balls in $2\times 2\times 2$ supercell including 96 atoms.}
\label{fig8}
\end{figure}
Finally we estimated the activation barrier for alkali atom diffusion mediated by M vacancy, which occurs through the exchange positions between the solute atom A and M vacancy.
Figure~\ref{fig8} shows the calculated activation barrier with depiction of migration path.
It turned out that in \ce{FeSn2} the activation barriers (0.17 and 0.44 eV for Na and Li) are lower than those in \ce{CoSn2} (0.83 and 0.88 eV).
Also Li diffusion occurs with lower activation barrier than Na diffusion.

\section{Conclusions}
In conclusion, we have performed first-principles calculations to study the effect of M/A substitution on structural, mechanical, lattice vibrational and electronic properties of intermetallic compounds \ce{MSn2} (M = Fe, Co; A = Li, Na).
The main conclusions can be summarized as follows:

\begin{enumerate}
\item M/A substitution leads to lattice expansion along the $a$- and $c$-axis. With this substitution, tetragonal ratio and unit cell volume also increase and thereby mass density decreases. Na substitution has more announced effect than Li. \ce{FeSn2} has less volume expansion than \ce{CoSn2}.

\item The calculated elastic constants of \ce{MSn2} and \ce{MASn2} satisfy the mechanical stability criteria for tetragonal crystal. M/A substitution reduces the mechanical stability and deformation resistance of \ce{MSn2}, which is more announced with Na substitution. The Pugh's and Poisson's ratios decrease with M/A substitution for Fe in \ce{FeSn2}, leading to its phase change from ductile to brittle, but increase for Co in \ce{CoSn2}, keeping its ductility.

\item Phonon dispersion curve of \ce{FeSn2} in primitive unit cell exhibits imaginary frequency (soft) mode around $\Gamma$ point, which might lead to phase transition from $I4/mcm$ to $I4/m$, whereas \ce{CoSn2} has all the real phonon frequencies. With M/A substitution for Fe, the soft mode of \ce{FeSn2} disappears for \ce{FeLiSn2} but is still preserved for \ce{FeNaSn2}. In phonon DOS, the gap between low and high frequency regions appears with this substitution.

\item In electronic band structures, there are two or four bands crossing the Fermi level $E_{\text{F}}$. For \ce{FeSn2}, the hole bands below $E_{\text{F}}$ is contributed equally from Fe-$3d$ and Sn-$5s, 5p$ states and the electron band above $E_{\text{F}}$ is dominated by Fe-$3d$ states. In contrast, the hole bands are dominated by Co-$3d$ states while the electron bands are contributed equally from Co-$3d$ and Sn-$5p$ states.

\item Formation enthalpy of \ce{FeSn2} compound with Fe vacancy or substitutional solute alkali atom is higher than that of \ce{CoSn2}. Li substitutional defect has lower formation energy than Na one. The activation barrier of vacancy-mediated Li diffusion is higher than Na one. The activation barriers in \ce{FeSn2} are lower than those in \ce{CoSn2}.

\end{enumerate}

\section*{Acknowledgments}
Computations in this work have been performed on the HP Blade System C7000 (HP BL460c) managed by Faculty of Materials Science, Kim Il Sung University.

\section*{\label{note}Notes}
The authors declare no competing financial interest.

\bibliographystyle{elsarticle-num-names}
\bibliography{Reference}

\begin{thebibliography}{42}
\providecommand{\natexlab}[1]{#1}
\providecommand{\url}[1]{\texttt{#1}}
\providecommand{\urlprefix}{URL }
\expandafter\ifx\csname urlstyle\endcsname\relax
  \providecommand{\doi}[1]{doi:\discretionary{}{}{}#1}\else
  \providecommand{\doi}[1]{doi:\discretionary{}{}{}\begingroup
  \urlstyle{rm}\url{#1}\endgroup}\fi
\providecommand{\bibinfo}[2]{#2}

\bibitem[{Vaalma et~al.(2018)Vaalma, Buchholz, Weil, and
  Passerini}]{Vaalma18nrm}
\bibinfo{author}{C.~Vaalma}, \bibinfo{author}{D.~Buchholz},
  \bibinfo{author}{M.~Weil}, \bibinfo{author}{S.~Passerini}, \bibinfo{title}{A
  cost and resource analysis of sodium-ion batteries}, \bibinfo{journal}{Nat.
  Rev. Mater.} \bibinfo{volume}{3} (\bibinfo{year}{2018})
  \bibinfo{pages}{18013}.

\bibitem[{Mukherjee et~al.(2019)Mukherjee, Mujib, Soares, and
  Singh}]{Mukherjee19m}
\bibinfo{author}{S.~Mukherjee}, \bibinfo{author}{S.~B. Mujib},
  \bibinfo{author}{D.~Soares}, \bibinfo{author}{G.~Singh},
  \bibinfo{title}{Electrode materials for high-performance sodium-ion
  batteries}, \bibinfo{journal}{Materials} \bibinfo{volume}{12}
  (\bibinfo{year}{2019}) \bibinfo{pages}{1952}.

\bibitem[{Bai et~al.(2017)Bai, Yang, Chen, and Mo}]{Bai17aem}
\bibinfo{author}{Q.~Bai}, \bibinfo{author}{L.~Yang}, \bibinfo{author}{H.~Chen},
  \bibinfo{author}{Y.~Mo}, \bibinfo{title}{Computational Studies of Electrode
  Materials in Sodium-Ion Batteries}, \bibinfo{journal}{Adv. Energy Mater.}
  \bibinfo{volume}{8} (\bibinfo{year}{2017}) \bibinfo{pages}{1702998}.

\bibitem[{Luo et~al.(2016)Luo, Shen, Bommier, Zhu, Ji, and Hu}]{Luo16acr}
\bibinfo{author}{W.~Luo}, \bibinfo{author}{F.~Shen},
  \bibinfo{author}{C.~Bommier}, \bibinfo{author}{H.~Zhu},
  \bibinfo{author}{X.~Ji}, \bibinfo{author}{L.~Hu}, \bibinfo{title}{Na-ion
  battery anodes: materials and electrochemistry}, \bibinfo{journal}{Acc. Chem.
  Res.} \bibinfo{volume}{49} (\bibinfo{year}{2016}) \bibinfo{pages}{231--240}.

\bibitem[{Li et~al.(2018)Li, Zheng, Zhang, Yang, Shao, and Guo}]{Li18ees}
\bibinfo{author}{L.~Li}, \bibinfo{author}{Y.~Zheng}, \bibinfo{author}{S.~L.
  Zhang}, \bibinfo{author}{J.~P. Yang}, \bibinfo{author}{Z.~P. Shao},
  \bibinfo{author}{Z.~P. Guo}, \bibinfo{title}{Recent progress on sodium ion
  batteries: potential high-performance anodes}, \bibinfo{journal}{Energy
  Environ. Sci.} \bibinfo{volume}{11} (\bibinfo{year}{2018})
  \bibinfo{pages}{2310--2340}.

\bibitem[{Huang et~al.(2018)Huang, Pan, Su, and An}]{Huang18jps}
\bibinfo{author}{B.~Huang}, \bibinfo{author}{Z.~Pan}, \bibinfo{author}{X.~Su},
  \bibinfo{author}{L.~An}, \bibinfo{title}{Tin-based materials as versatile
  anodes for alkali (earth)-ion batteries}, \bibinfo{journal}{J. Power Sources}
  \bibinfo{volume}{395} (\bibinfo{year}{2018}) \bibinfo{pages}{41--59}.

\bibitem[{Stratford et~al.(2017)Stratford, Mayo, Allan, Pecher, Borkiewicz,
  Wiaderek, Chapman, Pickard, Morris, and Grey}]{Stratford17jacs}
\bibinfo{author}{J.~M. Stratford}, \bibinfo{author}{M.~Mayo},
  \bibinfo{author}{P.~K. Allan}, \bibinfo{author}{O.~Pecher},
  \bibinfo{author}{O.~J. Borkiewicz}, \bibinfo{author}{K.~M. Wiaderek},
  \bibinfo{author}{K.~W. Chapman}, \bibinfo{author}{C.~J. Pickard},
  \bibinfo{author}{A.~J. Morris}, \bibinfo{author}{C.~P. Grey},
  \bibinfo{title}{Investigating sodium storage mechanisms in tin anodes: A
  combined pair distribution function analysis, density functional theory, and
  solid-state NMR approach}, \bibinfo{journal}{J. Am. Chem. Soc.}
  \bibinfo{volume}{139} (\bibinfo{year}{2017}) \bibinfo{pages}{7273--7286}.

\bibitem[{Li et~al.(2015)Li, Ding, and Mitlin}]{Li15acr}
\bibinfo{author}{Z.~Li}, \bibinfo{author}{J.~Ding},
  \bibinfo{author}{D.~Mitlin}, \bibinfo{title}{Tin and tin compounds for sodium
  ion battery anodes: Phase transformations and performance},
  \bibinfo{journal}{Acc. Chem. Res.} \bibinfo{volume}{48}
  (\bibinfo{year}{2015}) \bibinfo{pages}{1657--1665}.

\bibitem[{Baggetto et~al.(2014)Baggetto, Bridges, Jumas, Mullins, Carroll,
  Meisner, Crumlin, Liu, Yang, and Veith}]{Baggetto14jmca}
\bibinfo{author}{L.~Baggetto}, \bibinfo{author}{C.~A. Bridges},
  \bibinfo{author}{J.-C. Jumas}, \bibinfo{author}{D.~R. Mullins},
  \bibinfo{author}{K.~J. Carroll}, \bibinfo{author}{R.~A. Meisner},
  \bibinfo{author}{E.~J. Crumlin}, \bibinfo{author}{X.~Liu},
  \bibinfo{author}{W.~Yang}, \bibinfo{author}{G.~M. Veith}, \bibinfo{title}{The
  local atomic structure and chemical bonding in sodium tin phases},
  \bibinfo{journal}{J. Mater. Chem. A} \bibinfo{volume}{2}
  (\bibinfo{year}{2014}) \bibinfo{pages}{18959--18973}.

\bibitem[{Wang et~al.(2012)Wang, Liu, Mao, and Huang}]{Wang12nl}
\bibinfo{author}{J.~W. Wang}, \bibinfo{author}{X.~H. Liu},
  \bibinfo{author}{S.~X. Mao}, \bibinfo{author}{J.~Y. Huang},
  \bibinfo{title}{Microstructural evolution of tin nanoparticles during in situ
  sodium insertion and extraction}, \bibinfo{journal}{Nano Lett.}
  \bibinfo{volume}{12} (\bibinfo{year}{2012}) \bibinfo{pages}{5897--5902}.

\bibitem[{Wang et~al.(2019)Wang, Zhang, Li, and Shi}]{Wang19jac}
\bibinfo{author}{W.~Wang}, \bibinfo{author}{J.~Zhang}, \bibinfo{author}{B.~Li},
  \bibinfo{author}{L.~Shi}, \bibinfo{title}{Electrochemical investigation of
  Sn-Co alloys as anode for Na-ion batteries}, \bibinfo{journal}{J. Alloy
  Compd.} \bibinfo{volume}{780} (\bibinfo{year}{2019})
  \bibinfo{pages}{565--569}.

\bibitem[{Walter et~al.(2016)Walter, Doswald, and Kovalenko}]{Walter16jmca}
\bibinfo{author}{M.~Walter}, \bibinfo{author}{S.~Doswald},
  \bibinfo{author}{M.~V. Kovalenko}, \bibinfo{title}{Inexpensive colloidal SnSb
  nanoalloys as efficient anode materials for lithium- and sodium-ion
  batteries}, \bibinfo{journal}{J. Mater. Chem. A} \bibinfo{volume}{4}
  (\bibinfo{year}{2016}) \bibinfo{pages}{7053--7059}.

\bibitem[{Zhang et~al.(2017)Zhang, Wang, Ma, Yu, Lu, and Liu}]{Zhang17ra}
\bibinfo{author}{R.~Zhang}, \bibinfo{author}{Z.~Wang}, \bibinfo{author}{W.~Ma},
  \bibinfo{author}{W.~Yu}, \bibinfo{author}{S.~Lu}, \bibinfo{author}{X.~Liu},
  \bibinfo{title}{Improved sodium-ion storage properties by fabricating
  nanoporous CuSn alloy architecture}, \bibinfo{journal}{RSC Adv.}
  \bibinfo{volume}{7} (\bibinfo{year}{2017}) \bibinfo{pages}{29458--29463}.

\bibitem[{Wang et~al.(2010)Wang, Han, Chen, and Graetz}]{Wang10aami}
\bibinfo{author}{X.-L. Wang}, \bibinfo{author}{W.-Q. Han},
  \bibinfo{author}{J.~Chen}, \bibinfo{author}{J.~Graetz},
  \bibinfo{title}{Single-crystal intermetallic M-Sn (M = Fe, Cu, Co, Ni)
  nanospheres as negative electrodes for lithium-ion batteries},
  \bibinfo{journal}{ACS Appl. Mater. Int.} \bibinfo{volume}{2}
  (\bibinfo{year}{2010}) \bibinfo{pages}{1548--1551}.

\bibitem[{Chamas et~al.(2013)Chamas, Sougrati, Reibel, and
  Lippens}]{Chamas13cm}
\bibinfo{author}{M.~Chamas}, \bibinfo{author}{M.-T. Sougrati},
  \bibinfo{author}{C.~Reibel}, \bibinfo{author}{P.-E. Lippens},
  \bibinfo{title}{Quantitative analysis of the initial restructuring step of
  nanostructured FeSn$_2$‑based anodes for Li-ion batteries},
  \bibinfo{journal}{Chem. Mater.} \bibinfo{volume}{25} (\bibinfo{year}{2013})
  \bibinfo{pages}{2410--2420}.

\bibitem[{Wang et~al.(2018)Wang, He, Walter, Krumeich, Kravchyk, and
  Kovalenko}]{Wang18n}
\bibinfo{author}{S.~Wang}, \bibinfo{author}{M.~He},
  \bibinfo{author}{M.~Walter}, \bibinfo{author}{F.~Krumeich},
  \bibinfo{author}{K.~V. Kravchyk}, \bibinfo{author}{M.~V. Kovalenko},
  \bibinfo{title}{Monodisperse CoSn$_2$ and FeSn$_2$ nano crystals as
  high-performance anode materials for lithium-ion batteries},
  \bibinfo{journal}{Nanoscale} \bibinfo{volume}{10} (\bibinfo{year}{2018})
  \bibinfo{pages}{6827--6831}.

\bibitem[{Vogt and Villevieille(2017)}]{Vogt17jmca}
\bibinfo{author}{L.~O. Vogt}, \bibinfo{author}{C.~Villevieille},
  \bibinfo{title}{Elucidation of the reaction mechanisms of isostructural
  FeSn$_2$ and CoSn$_2$ negative electrodes for Na-ion batteries},
  \bibinfo{journal}{J. Mater. Chem. A} \bibinfo{volume}{5}
  (\bibinfo{year}{2017}) \bibinfo{pages}{3865--3874}.

\bibitem[{Vogt and Villevieille(2016)}]{Vogt16jes}
\bibinfo{author}{L.~O. Vogt}, \bibinfo{author}{C.~Villevieille},
  \bibinfo{title}{FeSn$_2$ and CoSn$_2$ electrode materials for Na-ion
  batteries}, \bibinfo{journal}{J. Electrochem. Soc.} \bibinfo{volume}{163}
  (\bibinfo{year}{2016}) \bibinfo{pages}{A1306--A1310}.

\bibitem[{Yui et~al.(2015)Yui, Ono, Hayashi, Nemoto, Hayashi, Asakura, and
  Kitabayashi}]{Yui15jes}
\bibinfo{author}{Y.~Yui}, \bibinfo{author}{Y.~Ono},
  \bibinfo{author}{M.~Hayashi}, \bibinfo{author}{Y.~Nemoto},
  \bibinfo{author}{K.~Hayashi}, \bibinfo{author}{K.~Asakura},
  \bibinfo{author}{H.~Kitabayashi}, \bibinfo{title}{Sodium-ion
  insertion/extraction properties of Sn-Co anodes and Na pre-doped Sn-Co
  anodes}, \bibinfo{journal}{J. Electrochem. Soc.} \bibinfo{volume}{162}
  (\bibinfo{year}{2015}) \bibinfo{pages}{A3098--A3102}.

\bibitem[{Gonzalez et~al.(2013)Gonzalez, Nacimiento, Alcantara, Ortiz, and
  Tirado}]{Gonzalez13cec}
\bibinfo{author}{J.~R. Gonzalez}, \bibinfo{author}{F.~Nacimiento},
  \bibinfo{author}{R.~Alcantara}, \bibinfo{author}{G.~F. Ortiz},
  \bibinfo{author}{J.~L. Tirado}, \bibinfo{title}{Electrodeposited CoSn$_2$ on
  nickel open-cell foam: advancing toward high power lithium ion and sodium ion
  batteries}, \bibinfo{journal}{CrystEngComm} \bibinfo{volume}{15}
  (\bibinfo{year}{2013}) \bibinfo{pages}{9196--9202}.

\bibitem[{Chamas et~al.(2011{\natexlab{a}})Chamas, Lippens, Jumas, Hassoun,
  Panero, and Scrosati}]{Chamas11ea}
\bibinfo{author}{M.~Chamas}, \bibinfo{author}{P.~E. Lippens},
  \bibinfo{author}{J.~C. Jumas}, \bibinfo{author}{J.~Hassoun},
  \bibinfo{author}{S.~Panero}, \bibinfo{author}{B.~Scrosati},
  \bibinfo{title}{Electrochemical impedance characterization of FeSn$_2$
  electrodes for Li-ion batteries}, \bibinfo{journal}{Electrochim. Acta}
  \bibinfo{volume}{56} (\bibinfo{year}{2011}{\natexlab{a}})
  \bibinfo{pages}{6732--6736}.

\bibitem[{Chamas et~al.(2011{\natexlab{b}})Chamas, Lippens, Jumas, Boukerma,
  Dedryvere, Gonbeau, Hassoun, Panero, and Scrosati}]{Chamas11jps}
\bibinfo{author}{M.~Chamas}, \bibinfo{author}{P.~E. Lippens},
  \bibinfo{author}{J.~C. Jumas}, \bibinfo{author}{K.~Boukerma},
  \bibinfo{author}{R.~Dedryvere}, \bibinfo{author}{D.~Gonbeau},
  \bibinfo{author}{J.~Hassoun}, \bibinfo{author}{S.~Panero},
  \bibinfo{author}{B.~Scrosati}, \bibinfo{title}{Comparison between
  microparticles and nanostructured particles of FeSn$_2$ as anode materials
  for Li-ion batteries}, \bibinfo{journal}{J. Power Sources}
  \bibinfo{volume}{196} (\bibinfo{year}{2011}{\natexlab{b}})
  \bibinfo{pages}{7011--7015}.

\bibitem[{Zhang et~al.(2008)Zhang, Tu, Huang, Yuan, Wang, and Mao}]{Zhang08jac}
\bibinfo{author}{C.~Q. Zhang}, \bibinfo{author}{J.~P. Tu},
  \bibinfo{author}{X.~H. Huang}, \bibinfo{author}{Y.~F. Yuan},
  \bibinfo{author}{S.~F. Wang}, \bibinfo{author}{F.~Mao},
  \bibinfo{title}{Preparation and electrochemical performances of nanoscale
  FeSn$_2$ as anode material for lithium ion batteries}, \bibinfo{journal}{J.
  Alloy Compd.} \bibinfo{volume}{457} (\bibinfo{year}{2008})
  \bibinfo{pages}{81--85}.

\bibitem[{Edison et~al.(2017{\natexlab{a}})Edison, Satish, Ling, Bucher,
  Aravindan, and Madhavi}]{Edison17jps}
\bibinfo{author}{E.~Edison}, \bibinfo{author}{R.~Satish},
  \bibinfo{author}{W.~C. Ling}, \bibinfo{author}{N.~Bucher},
  \bibinfo{author}{V.~Aravindan}, \bibinfo{author}{S.~Madhavi},
  \bibinfo{title}{Nanostructured intermetallic FeSn$_2$-carbonaceous composites
  as highly stable anode for Na-ion batteries}, \bibinfo{journal}{J. Power
  Sources} \bibinfo{volume}{343} (\bibinfo{year}{2017}{\natexlab{a}})
  \bibinfo{pages}{296--302}.

\bibitem[{Edison et~al.(2017{\natexlab{b}})Edison, Ling, Aravindan, and
  Madhavi}]{Edison17cec}
\bibinfo{author}{E.~Edison}, \bibinfo{author}{W.~C. Ling},
  \bibinfo{author}{V.~Aravindan}, \bibinfo{author}{S.~Madhavi},
  \bibinfo{title}{Highly stable intermetallic FeSn$_2$-graphite composite anode
  for sodium-ion batteries}, \bibinfo{journal}{ChemElectroChem}
  \bibinfo{volume}{4} (\bibinfo{year}{2017}{\natexlab{b}})
  \bibinfo{pages}{1932--1936}.

\bibitem[{Liu et~al.(2015)Liu, Xie, Zhao, Lv, Wang, Feng, and
  \'{S}wierczek}]{Liu15ssi}
\bibinfo{author}{X.~Liu}, \bibinfo{author}{J.~Xie}, \bibinfo{author}{H.~Zhao},
  \bibinfo{author}{P.~Lv}, \bibinfo{author}{K.~Wang},
  \bibinfo{author}{Z.~Feng}, \bibinfo{author}{K.~\'{S}wierczek},
  \bibinfo{title}{Electrochemical properties of mechanochemically synthesized
  CoSn$_2$-C nanocomposite-type anode material for Li-ion batteries},
  \bibinfo{journal}{Solid State Ionics} \bibinfo{volume}{269}
  (\bibinfo{year}{2015}) \bibinfo{pages}{86--92}.

\bibitem[{Nacimiento et~al.(2012)Nacimiento, Alc\'{a}ntara, Nwokeke,
  Gonz\'{a}lez, and Tirado}]{Nacimiento12us}
\bibinfo{author}{F.~Nacimiento}, \bibinfo{author}{R.~Alc\'{a}ntara},
  \bibinfo{author}{U.~G. Nwokeke}, \bibinfo{author}{J.~R. Gonz\'{a}lez},
  \bibinfo{author}{J.~L. Tirado}, \bibinfo{title}{Nanocrystalline
  CoSn$_2$-carbon composite electrode prepared by using sonochemistry},
  \bibinfo{journal}{Ultrasonics Sonochem.} \bibinfo{volume}{19}
  (\bibinfo{year}{2012}) \bibinfo{pages}{352--357}.

\bibitem[{Leibowitz et~al.(2015)Leibowitz, Allcorn, and
  Manthiram}]{Leibowitz15jps}
\bibinfo{author}{J.~Leibowitz}, \bibinfo{author}{E.~Allcorn},
  \bibinfo{author}{A.~Manthiram}, \bibinfo{title}{FeSn$_2$-TiC nanocomposite
  alloy anodes for lithium ion batteries}, \bibinfo{journal}{J. Power Sources}
  \bibinfo{volume}{295} (\bibinfo{year}{2015}) \bibinfo{pages}{125--130}.

\bibitem[{Guo et~al.(2007)Guo, Zhao, Jia, Li, and Qiu}]{Guo07ea}
\bibinfo{author}{H.~Guo}, \bibinfo{author}{H.~Zhao}, \bibinfo{author}{X.~Jia},
  \bibinfo{author}{X.~Li}, \bibinfo{author}{W.~Qiu}, \bibinfo{title}{A novel
  micro-spherical CoSn$_2$/Sn alloy composite as high capacity anode materials
  for Li-ion rechargeable batteries}, \bibinfo{journal}{Electrochim. Acta}
  \bibinfo{volume}{52} (\bibinfo{year}{2007}) \bibinfo{pages}{4853--4857}.

\bibitem[{Sun et~al.(2016)Sun, Zhang, Liu, Wang, and Bu}]{Sun16cms}
\bibinfo{author}{W.~Sun}, \bibinfo{author}{L.~Zhang}, \bibinfo{author}{J.~Liu},
  \bibinfo{author}{H.~Wang}, \bibinfo{author}{Y.~Bu},
  \bibinfo{title}{First-principles investigation of mechanical, thermodynamic
  and electronic properties of FeSn$_5$ and CoSn$_5$ phases},
  \bibinfo{journal}{Comput. Mater. Sci.} \bibinfo{volume}{111}
  (\bibinfo{year}{2016}) \bibinfo{pages}{175--180}.

\bibitem[{{de Jong} et~al.(2015){de Jong}, Chen, Angsten, Jain, Notestine, and
  Gamst}]{Jong15sd}
\bibinfo{author}{M.~{de Jong}}, \bibinfo{author}{W.~Chen},
  \bibinfo{author}{T.~Angsten}, \bibinfo{author}{A.~Jain},
  \bibinfo{author}{R.~Notestine}, \bibinfo{author}{A.~Gamst},
  \bibinfo{title}{Charting the complete elastic properties of inorganic
  crystalline compounds}, \bibinfo{journal}{Sci. Data} \bibinfo{volume}{2}
  (\bibinfo{year}{2015}) \bibinfo{pages}{150009}.

\bibitem[{Armbr\"{u}ster et~al.(2007)Armbr\"{u}ster, Schmidt, Cardoso-Gil,
  Borrmann, and Grin}]{Armbruster07zkn}
\bibinfo{author}{M.~Armbr\"{u}ster}, \bibinfo{author}{M.~Schmidt},
  \bibinfo{author}{R.~Cardoso-Gil}, \bibinfo{author}{H.~Borrmann},
  \bibinfo{author}{Y.~Grin}, \bibinfo{title}{Crystal structures of iron
  distannide, FeSn$_2$, and cobalt distannide, CoSn$_2$}, \bibinfo{journal}{Z.
  Kristallogr. NCS} \bibinfo{volume}{222} (\bibinfo{year}{2007})
  \bibinfo{pages}{83--84}.

\bibitem[{{P. Giannozzi, S. Baroni, N. Bonini, M. Calandra, R. Car, et
  al.}(2009)}]{QE09jpcm}
\bibinfo{author}{{P. Giannozzi, S. Baroni, N. Bonini, M. Calandra, R. Car, et
  al.}}, \bibinfo{title}{QUANTUM ESPRESSO: A modular and open-source software
  project for quantum simulations of materials}, \bibinfo{journal}{J. Phys.:
  Condens. Matter} \bibinfo{volume}{21} (\bibinfo{year}{2009})
  \bibinfo{pages}{395502}.

\bibitem[{Perdew et~al.(1996)Perdew, Burke, and Ernzerhof}]{PBE96prl}
\bibinfo{author}{J.~P. Perdew}, \bibinfo{author}{K.~Burke},
  \bibinfo{author}{M.~Ernzerhof}, \bibinfo{title}{Generalized gradient
  approximation made simple}, \bibinfo{journal}{Phys. Rev. Lett.}
  \bibinfo{volume}{77} (\bibinfo{year}{1996}) \bibinfo{pages}{3865}.

\bibitem[{Togo et~al.(2008)Togo, Oba, and Tanaka}]{phonopy}
\bibinfo{author}{A.~Togo}, \bibinfo{author}{F.~Oba},
  \bibinfo{author}{I.~Tanaka}, \bibinfo{title}{First-principles calculations of
  the ferroelastic transition between rutile-type and \ce{CaCl2}-type \ce{SiO2}
  at high pressures}, \bibinfo{journal}{Phys. Rev. B} \bibinfo{volume}{78}
  (\bibinfo{year}{2008}) \bibinfo{pages}{134106}.

\bibitem[{Henkelman et~al.(2000)Henkelman, Uberuaga, and
  J\'{o}nsson}]{NEB00jcp}
\bibinfo{author}{G.~Henkelman}, \bibinfo{author}{B.~P. Uberuaga},
  \bibinfo{author}{H.~J\'{o}nsson}, \bibinfo{title}{A climbing image nudged
  elastic band method for finding saddle points and minimum energy paths},
  \bibinfo{journal}{J. Chem. Phys.} \bibinfo{volume}{113}
  (\bibinfo{year}{2000}) \bibinfo{pages}{9901--9904}.

\bibitem[{Momma and Izumi(2011)}]{VESTA11jac}
\bibinfo{author}{K.~Momma}, \bibinfo{author}{F.~Izumi}, \bibinfo{title}{VESTA 3
  for three-dimensional visualization of crystal, volumetric and morphology
  data}, \bibinfo{journal}{J. Appl. Crystallogr.} \bibinfo{volume}{44}
  (\bibinfo{year}{2011}) \bibinfo{pages}{1272--1276}.

\bibitem[{Max and Kun(1956)}]{Max56}
\bibinfo{author}{B.~Max}, \bibinfo{author}{H.~Kun}, \bibinfo{title}{Dynamical
  theory of crystal lattices}, \bibinfo{publisher}{Clarendon, Oxford, UK},
  \bibinfo{year}{1956}.

\bibitem[{Chiodo et~al.(2006)Chiodo, Gotsis, Russo, and Sicilia}]{Chiodo06cpl}
\bibinfo{author}{S.~Chiodo}, \bibinfo{author}{H.~J. Gotsis},
  \bibinfo{author}{N.~Russo}, \bibinfo{author}{E.~Sicilia},
  \bibinfo{title}{OsB$_2$ and RuB$_2$, ultra incompressible, hard materials:
  First-principles electronic structure calculations}, \bibinfo{journal}{Chem.
  Phys. Lett.} \bibinfo{volume}{425} (\bibinfo{year}{2006})
  \bibinfo{pages}{311--314}.

\bibitem[{Pugh(1954)}]{Pugh54pm}
\bibinfo{author}{S.~F. Pugh}, \bibinfo{title}{XCII. Relations between the
  elastic moduli and the plastic properties of polycrystalline pure metals},
  \bibinfo{journal}{Phil. Mag.} \bibinfo{volume}{45} (\bibinfo{year}{1954})
  \bibinfo{pages}{823--843}.

\bibitem[{Hadi et~al.(2017)Hadi, Roknuzzaman, Chroneos, Naqib, Islam, Vovk, and
  Ostrikov}]{Hadi17cms}
\bibinfo{author}{M.~A. Hadi}, \bibinfo{author}{M.~Roknuzzaman},
  \bibinfo{author}{A.~Chroneos}, \bibinfo{author}{S.~H. Naqib},
  \bibinfo{author}{A.~K. M.~A. Islam}, \bibinfo{author}{R.~V. Vovk},
  \bibinfo{author}{K.~Ostrikov}, \bibinfo{title}{Elastic and thermodynamic
  properties of new (Zr$_{3-x}$Ti$_x$)AlC$_2$ MAX-phase solid solutions},
  \bibinfo{journal}{Comput. Mater. Sci.} \bibinfo{volume}{137}
  (\bibinfo{year}{2017}) \bibinfo{pages}{318--326}.

\bibitem[{Zhou et~al.(2014)Zhou, Wang, Cui, and Li}]{Zhou14jap}
\bibinfo{author}{D.~Zhou}, \bibinfo{author}{J.~Wang}, \bibinfo{author}{Q.~Cui},
  \bibinfo{author}{Q.~Li}, \bibinfo{title}{Crystal structure and physical
  properties of Mo$_2$B: First-principles calculations}, \bibinfo{journal}{J.
  Appl. Phys.} \bibinfo{volume}{115} (\bibinfo{year}{2014})
  \bibinfo{pages}{113504--113508}.

\end{thebibliography}

\end{document}